\documentclass[a4paper,11pt]{article}
\pdfoutput=1
\usepackage{jheppub}
\usepackage[T1]{fontenc}
\usepackage{graphicx}
\usepackage{amsmath}
\usepackage{amsmath}
\usepackage{xspace}
\usepackage{subfig}
\usepackage{xcolor}
\usepackage{mathtools}
\usepackage{braket}
\usepackage{multirow}

\newcommand\vecl[1]{{\bf{#1}}}        			  		 	 
\newcommand\vecs[1]{\boldsymbol{#1}}    				   	 
\newcommand\nn[0]{\nonumber}					 			 
\newcommand\vvbb[0]{2\nu\beta\beta}			 			 

\title{Two-Neutrino Double Beta Decay\\ with Sterile Neutrinos}

\author[a]{Patrick D. Bolton} 
\emailAdd{patrick.bolton.17@ucl.ac.uk}
\author[a]{Frank F. Deppisch} 
\emailAdd{f.deppisch@ucl.ac.uk}\affiliation[a]{\AddrUCL}
\author[b]{Luk\'{a}\v{s} Gr\'{a}f}
\emailAdd{lukas.graf@mpi-hd.mpg.de}\affiliation[b]{\AddrMPIK}
\author[c]{Fedor {\v S}imkovic}
\emailAdd{fedor.simkovic@fmph.uniba.sk}\affiliation[c]{\AddrBLTP}\affiliation{\AddrComenius}\affiliation{\AddrIEAP}

\newcommand{\AddrUCL}{Department of Physics and Astronomy, University College London,\\London WC1E 6BT, United Kingdom}

\newcommand{\AddrMPIK}{Max-Planck-Institut f\"ur Kernphysik, Saupfercheckweg 1, 69117 Heidelberg, Germany}

\newcommand{\AddrBLTP}{BLTP, JINR, 141980 Dubna, Russia}

\newcommand{\AddrComenius}{Comenius University, Mlynsk\'a dolina F1, SK–842 48 Bratislava, Slovakia}

\newcommand{\AddrIEAP}{IEAP CTU, 128–00 Prague, Czech Republic}

\abstract{Usually considered a background for experimental searches for the hypothetical neutrinoless double beta decay process, two-neutrino double beta decay nevertheless provides a complementary probe of physics beyond the Standard Model. In this paper we investigate how the presence of a sterile neutrino, coupled to the Standard Model either via a left-handed or right-handed current, affects the energy distribution and angular correlation of the outgoing electrons in two-neutrino double beta decay. We pay particular attention on the behaviour of the energy distribution at the kinematic endpoint and we estimate the current limits on the active-sterile mixing and effective right-handed coupling using current experimental data as a function of the sterile neutrino mass. We also investigate the sensitivities of future experiments. Our results complement the corresponding constraints on sterile neutrinos from single beta decay measurements in the 0.1 -- 10 MeV mass range.}

\begin{document}	
\maketitle
\flushbottom


\section{Introduction}

Sterile neutrinos are among the most sought-after candidates of exotic particles. The main motivation for their existence is the fact that the Standard Model (SM) does not contain right-handed (RH) counterparts of the left-handed neutrino states participating in electroweak interactions, in contrast to the quarks and charged leptons. Their absence is purely because they are required to be singlets under the weak $SU(2)_L$ and have zero weak hypercharge if they are to participate in a Yukawa interaction with the left-handed neutrino states and the SM Higgs. They are thus truly sterile with respect to the SM gauge group and in this paradigm can only manifest themselves through an admixture with the active neutrinos. Thus, sterile neutrinos can in fact be considered to be any exotic fermion that is uncharged under the SM gauge interactions; unless protected by some new symmetry they will mix with the active neutrinos as described above. Hence, sterile neutrinos are also often referred to as heavy neutral leptons.

An important feature of their mixing with the active neutrinos is the resulting impact on the light neutrino masses. Neutrino oscillations \cite{Tanabashi:2018oca} imply that the active neutrinos have small but non-zero masses. By adding a SM-singlet, RH neutrino field $\nu_R$ per generation to the SM, neutrinos can become massive. A so-called Dirac mass term can be generated through the Yukawa interaction with the SM Higgs, though the coupling required is tiny with effects unmeasurable experimentally. In any case, the sterile states will be allowed to acquire a so-called Majorana mass, unless protected by lepton number conserving symmetry, modifying the spectrum and nature of neutrinos considerably. This of course refers to the well-known type I seesaw mechanism~\cite{Minkowski:1977sc, Mohapatra:1979ia, Gellmann:1980vs, Yanagida:1979as, Schechter:1980gr}. The sterile neutrinos were initially considered to be very heavy ($m_{N}\sim 10^{14}$~GeV) in order to generate the correct light neutrino masses, but there is now a strong theoretical and experimental incentive to consider sterile neutrinos at accessible energies. Fig.~\ref{fig:current_constraints} summarises the current constraints on the active-sterile mixing strength $|V_{eN}|^2$ in the regime $1~\text{eV} < m_N < 10$~TeV derived from numerous experiments. The most stringent limits from fixed target and collider experiments can be found in the mass range $1~\text{GeV} < m_N < 100$ GeV, a region motivated by leptogenesis models.

Lighter sterile neutrino masses, while challenging to accommodate due to the constraints from astrophysics, are still of interest, especially around $m_N \sim 10$~keV where sterile neutrinos may act as warm dark matter. In the regime $10~\text{eV} < m_N < 1$ MeV, nuclear beta decays are currently the only laboratory-based experimental method able to probe sterile neutrinos. Neutrinoless double beta ($0\nu\beta\beta$) decay is an exception, setting stringent limits over the whole range in Fig.~\ref{fig:current_constraints}, but only if the sterile neutrinos are Majorana fermions -- for sterile Dirac neutrinos or quasi-Dirac neutrinos with relative splittings $\Delta m_N /m_N \lesssim 10^{-4}$ \cite{Deppisch:2020ztt}, the constraints vanish or become weaker respectively. In addition, if the sterile neutrinos in question are wholly responsible for giving mass to the active neutrinos via the type I seesaw mechanism, the contributions from active and sterile neutrinos to $0\nu\beta\beta$ decay cancel each other, see Sec.~\ref{sec:0vbb}. Thus, especially around $m_N\sim 1$~MeV, the current constraints are rather weak, of the order $|V_{eN}|^2 \lesssim$~few~$\times 10^{-3}$. As mentioned above, the constraints arise from searches for kinks in the electron energy spectrum and measurements of the $ft$ value of various beta decay isotopes, see Sec.~\ref{sec:beta-decay} for a brief review. 

This weakening of limits motivates the use of novel methods to constrain the active-sterile mixing in this mass regime. In this work we assess the potential of $0\nu\beta\beta$ decay experiments being sensitive to kinks in the background two-neutrino double beta ($2\nu\beta\beta$) decay spectrum caused by the presence of sterile neutrinos  in the final state with masses $m_N \lesssim 1$~MeV. This is fully analogous to the corresponding searches in single beta decays but $2\nu\beta\beta$ decaying isotopes typically have $Q$ values of a few MeV and are thus expected probe sterile neutrinos in such a mass range. The $2\nu\beta\beta$ decay process is of course very rare so it may at first seem difficult to achieve high enough statistics. While $2\nu\beta\beta$ decay is indeed not expected to improve the limits considerably, the $2\nu\beta\beta$ decays spectrum will be measured to high precision in several isotopes as $0\nu\beta\beta$ decay is searched for in ongoing and future experiments. The relevant data to look for sterile neutrinos in $2\nu\beta\beta$ decays will be available, which, generally speaking, can be used to look for signs of new physics in its own right \cite{Deppisch:2020mxv, Deppisch:2020sqh}.

In addition to a truly sterile neutrino, i.e. one that inherits the SM charged-current Fermi interaction albeit suppressed by the active-sterile mixing, we also consider RH current interactions of the \textquoteleft sterile' neutrino, e.g. arising in left-right symmetric models. Such interactions change the angular distribution of the electrons emitted in $2\nu\beta\beta$ decay \cite{Deppisch:2020mxv}. We parametrise all interactions in terms of effective operators of the SM with a light sterile neutrino, suitable in $2\nu\beta\beta$ decays with characteristic energies of $\lesssim 10$~MeV.

This paper is organised as follows. In Sec.~\ref{sec:model} we introduce the effective operators relevant for our discussions. In Sec.~\ref{sec:searches} we briefly review the current limits on the active-sterile mixing squared $|V_{eN}|^2$ with a focus on the mass regime $m_N \sim 1$~MeV. The calculation of the $2\nu\beta\beta$ decay spectrum with the emission of one sterile neutrino is described in Sec.~\ref{sec:dbd}. Sec.~\ref{sec:results} introduces our statistical procedure and presents the estimated current limits and prospective future sensitivities from sterile neutrino searches in $2\nu\beta\beta$ decay as our results. We conclude in Sec.~\ref{sec:conclusions}.

\section{Effective Interactions with Sterile Neutrinos}
\label{sec:model}

We consider the SM with the addition of a gauge singlet fermion $N$, i.e. the sterile neutrino. As we consider the second-order weak process of $2\nu\beta\beta$ decay, we restrict ourselves to the first generation of SM fermions. For processes with energies $\ll 100$~GeV we can describe the relevant weak processes using the effective SM Fermi interaction. The sterile neutrino inherits the Fermi interaction, but is suppressed by the active-sterile mixing $V_{eN}$. In addition, we allow the sterile neutrino to participate in exotic RH $V+A$ interactions. The effective Lagrangian taking into account the above takes the form
\begin{align}
\label{eq:lagrangian}
	\mathcal{L} &= 
	\frac{G_F\cos\theta_C}{\sqrt{2}}\left[
	  (1 + \delta_\text{SM}) j^\mu_L J^{\phantom{\mu}}_{L\mu} 
	     + V_{eN} j_L^{N\mu} J^{\phantom{\mu}}_{L\mu}
		 + \epsilon_{LR} j_R^{N\mu} J^{\phantom{\mu}}_{L\mu}
		 + \epsilon_{RR} j_R^{N\mu} J^{\phantom{\mu}}_{R\mu}
	\right] + \text{h.c.},
\end{align}
with the tree-level Fermi constant $G_F$, the Cabbibo angle $\theta_C$, and the leptonic and hadronic currents $j_L^\mu = \bar e\gamma^\mu(1-\gamma_5)\nu$, $j_{L,R}^{N\mu} = \bar e\gamma^\mu(1\mp\gamma_5)N$ and $J^\mu_{L,R} = \bar u\gamma^\mu(1\mp\gamma_5)d$, respectively. The SM electroweak radiative corrections are encoded in $\delta_{SM}$. The active-sterile mixing is $V_{eN}$ and the $\epsilon_{XY}$ encapsulate effects from integrating out new physics giving rise to $V+A$ currents of the sterile neutrino. We neglect any further effective operators, such as exotic contributions to the SM Fermi interaction and RH currents with the active neutrino \cite{Deppisch:2020mxv}.

In Eq.~\eqref{eq:lagrangian}, $\nu$ and $N$ are 4-spinor fields of the light electron neutrino and the sterile neutrino. They are either defined to be Majorana fermions, $\nu = \nu_L + \nu_L^c$, $N = N^{c}_R + N_R$ (i.e. a Majorana spinor constructed from the left-handed Weyl spinor and its charge-conjugate) or Dirac fermions $\nu = \nu_L + \nu_R$, $N = N_R + N_L$ (a Dirac spinor constructed from two different Weyl fields). The calculation of $2\nu\beta\beta$ decay is not affected by this, i.e. it is insensitive to the Dirac versus Majorana character. If the neutrinos are Majorana the constraints from $0\nu\beta\beta$ decay must be considered.


\section{Constraints on Sterile Neutrinos}
\label{sec:searches}

In this section we review the constraints on the active-sterile mixing strength squared $|V_{e N}|^2$ as a function of the sterile neutrino mass $m_N$. We mainly concentrate on limits in the $0.1 ~\text{MeV} < m_N < 3$~MeV mass range. This is because $2\nu\beta\beta$ decay measurements are only sensitive to sterile neutrino masses below the $Q$ value of the $2\nu\beta\beta$ decay process, which is of order $Q\sim 1 - 3$~MeV for the isotopes of interest. The relevant constraints in this range come from the non-observation of $0\nu\beta\beta$ decay, single beta decay spectra, sterile neutrino decays and cosmological probes. We will see that the same constraints also apply broadly to the RH current couplings $|\epsilon_{LR}|^2$ and $|\epsilon_{RR}|^2$.

\begin{figure}[t!]
	\centering
	\includegraphics[width=0.9\textwidth]{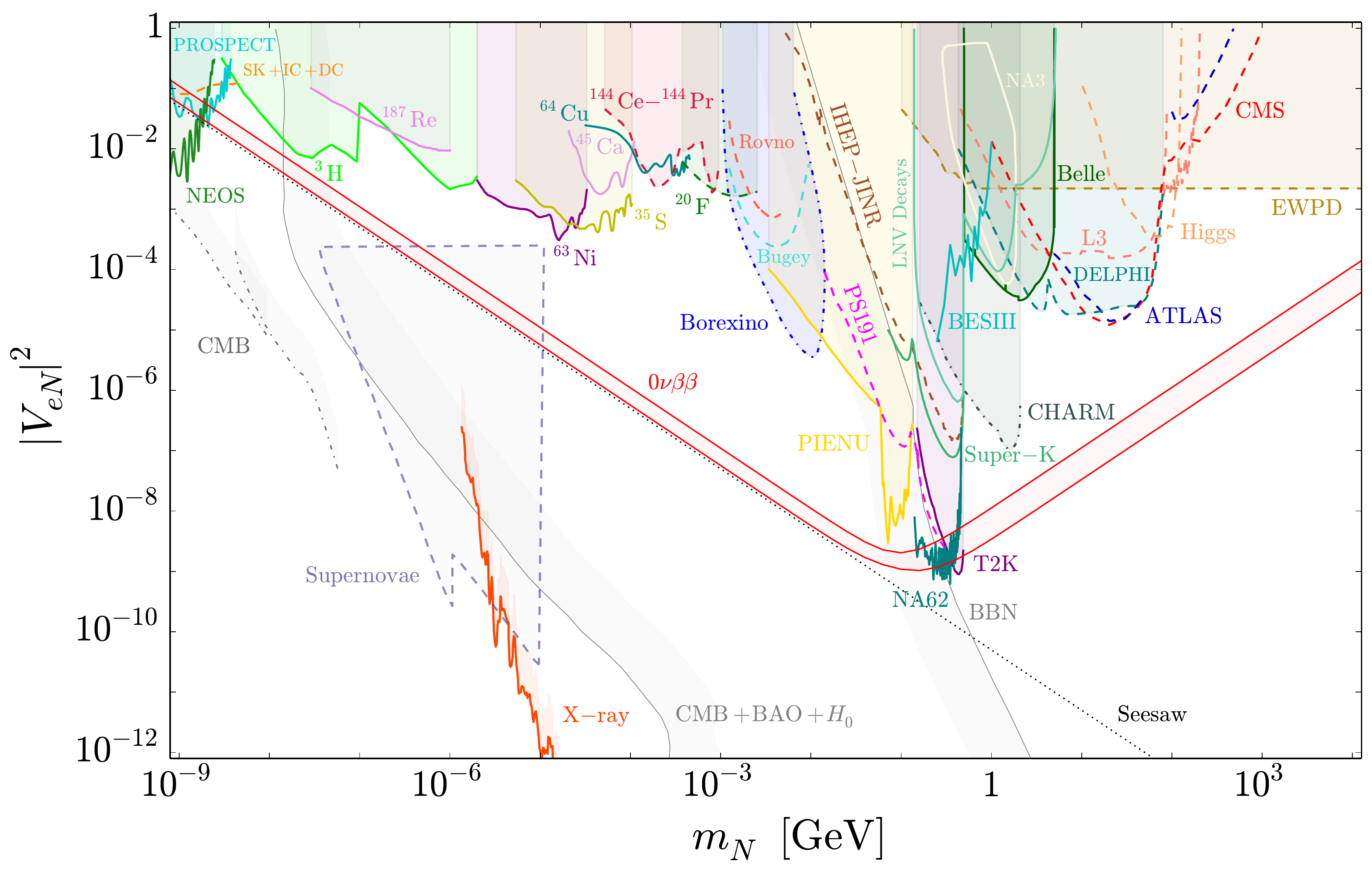}
	\caption{Constraints on the squared mixing strength $|V_{e N}|^2$ of the sterile neutrino with the electron neutrino as a function of its mass $m_N$. For simplicity we assume $\nu_e$ to be the only active neutrino. The shaded regions are excluded by the searches and observations as labelled. They are discussed in Sec.~4 of Ref.~\cite{Bolton:2019pcu}. The band labelled \textquoteleft $0\nu\beta\beta$' denotes the uncertainty on the current upper limit from $0\nu\beta\beta$ decay searches on a Majorana sterile neutrino. The diagonal black-dotted line labelled \textquoteleft Seesaw' indicates the canonical seesaw relation $|V_{eN}|^2 = m_{\nu_e}/m_N$ with $m_{\nu_e} = 0.05$~eV.}
	\label{fig:current_constraints}
\end{figure}

As an overview we show in Fig.~\ref{fig:current_constraints} the existing $|V_{e N}|^2$ constraints over the mass range $1~\text{eV} < m_N < 10$~TeV; for further information on each labelled constraint see Sec.~4 of Ref.~\cite{Bolton:2019pcu} and references therein. It is interesting to note the relative weakness of the upper limits from single beta decay experiments in the range $0.1~\text{MeV} < m_N < 3$~MeV. Mixing strengths are nonetheless excluded down to $|V_{eN}|^2 \lesssim 10^{-7} - 10^{-6}$ and $|V_{eN}|^2\lesssim 10^{-14}-10^{-11}$ by $0\nu\beta\beta$ decay and cosmological probes, respectively. It is crucial though to emphasise that the former constraints are model-dependent and can be avoided if neutrinos are Dirac fermions or if the sterile neutrinos are responsible for the light neutrino mass generation. Cosmological constraints rely on modelling of the early universe and can be avoided in extended scenarios where the sterile neutrinos have exotic interactions with a dark sector \cite{Bezrukov:2009th, Nemevsek:2012cd, El-Zant:2013nta, Biswas:2018iny}. This therefore motivates looking at the sensitivities of current and future $2\nu\beta\beta$ decay measurements but we first look at the existing constraints within the region of interest in more detail.

\subsection{Neutrinoless Double Beta Decay}
\label{sec:0vbb}
If we consider the active and sterile neutrinos to be purely Dirac fermions, lepton number is conserved and $0\nu\beta\beta$ decay is forbidden. Searches for this decay will thus not provide constraints on the active-sterile mixing of Dirac neutrinos. 

In the Majorana case, if $n_S$ sterile neutrinos are added to the SM with masses $m_{N_i}$ and active-sterile mixing strengths $V_{eN_{i}}$ (we assume for simplicity a single active state $\nu_e$), the inverse of the half-life $T_{1/2}^{0\nu}$ for the $0\nu\beta\beta$ decay process can be written using the interpolating formula
\begin{align}
	\frac{1}{T_{1/2}^{0 \nu}}=
	G^{0\nu} g_{A}^4|M^{0\nu}|^2
	\left|\frac{m_{\nu_e}}{m_{e}}
	+ \frac{ \left\langle\mathbf{p}^2\right\rangle}{m_e}\sum_{i=1}^{n_S} \frac{V_{e N_i}^2 m_{N_i}}{\left\langle\mathbf{p}^2\right\rangle + m_{N_i}^2}\right|^2.
\end{align}
Here $G^{0\nu}$ is the phase space factor, $g_{A}$ is the axial vector coupling, $M^{0\nu}$ is the light neutrino exchange nuclear matrix element and $\left\langle\mathbf{p}^2\right\rangle$ is the average momentum transfer of the process \cite{Kovalenko:2009td, Faessler:2014kka}. By considering a single sterile neutrino with mass $m_N$ and neglecting the contribution from the active neutrinos, the constraint in Fig.~\ref{fig:current_constraints} is derived using the current experimental bounds.

If the heavy states are related to the light state by a seesaw relation, then
\begin{align}
\label{eq:seesaw_rel}
	(\mathcal{M}_{\nu})_{11} = m_{\nu_{e}} + \sum_{i=1}^{n_S} V^2_{eN_i}m_{N_i} = 0\,,
\end{align}
must be satisfied. Thus, if the sterile states are lighter than the $0\nu\beta\beta$ decay momentum transfer, $m_{N_i}\ll \left\langle\mathbf{p}^2\right\rangle$, the $0\nu\beta\beta$ decay rate vanishes and the corresponding constraint in Fig.~\ref{fig:current_constraints} disappears. Sterile neutrinos have been discussed in the context of $0\nu\beta\beta$ decay in detail in Refs. \cite{Barea:2015zfa, Abada:2018qok, Bolton:2019pcu, Dekens:2020ttz}.

\subsection{Beta Decay}
\label{sec:beta-decay}
Electron neutrinos are produced in the beta decays of unstable isotopes via the LH charged-current interaction. If the active-sterile mixing strength $|V_{eN}|^2$ or RH couplings $|\epsilon_{LR}|^2$, $|\epsilon_{RR}|^2$ are non-zero, sterile neutrinos can be produced if their masses are smaller than the $Q$ value of the process. For a large enough $m_N$ the emission results in a distortion or \textquoteleft kink' in the beta decay spectrum and associated Kurie plot. 

The beta decay spectrum with respect to the kinetic energy of the emitted electron can be written for a single sterile neutrino with mixing as the incoherent sum
\begin{align}
\label{eq:betadecayspectrum}
	\frac{d\Gamma^\beta}{dE_e} =  \left(1-|V_{eN}|^2\right)\frac{d\Gamma^\nu(0)}{dE_e}
	+ |V_{e N}|^2 \frac{d\Gamma^\nu(m_N)}{dE_e},
\end{align}
where we neglect the light neutrino masses in the standard contribution. Due to unitarity, the contribution from the light neutrinos is reduced by the active-sterile mixing strength. The sterile neutrino contribution gives rise to a kink in the spectrum of relative size $|V_{eN}|^2$ and at electron energies $E_e = Q - m_N$. Alternatively, in the case the sterile neutrinos are produced by a RH current, the SM contribution is no longer reduced as a result of unitarity. This weakens the upper limits on $|\epsilon_{LR}|^2$ and $|\epsilon_{RR}|^2$ compared to $|V_{eN}|^2$, though the effect is negligible for upper bounds below $10^{-2}$.

Kink searches have been conducted for a variety of isotopes with different $Q$ values, making them sensitive to a range of sterile neutrino masses. Shown in Fig.~\ref{fig:current_constraints} are upper limits from the isotopes $^3$H~\cite{Hiddemann:1995ce, Kraus:2012he, Belesev:2013cba, Abdurashitov:2017kka}, $^{20}$F~\cite{PhysRevC.27.1175}, $^{35}$S~\cite{Holzschuh:2000nj},  $^{45}$Ca~\cite{Derbin:1997ut}, $^{63}$Ni~\cite{Holzschuh:1999vy}, $^{64}$Cu~\cite{Schreckenbach:1983cg}, $^{144}$Ce--$^{144}$Pr~\cite{Derbin2018} and  $^{187}$Re~\cite{PhysRevLett.86.1978}, assuming there to be a single sterile state. With smaller $Q$ values, $^{3}$H and $^{187}$Re provide constraints over the range $1~\text{eV} < m_N  < 1$~keV. It can be seen that $^{45}$Ca, $^{64}$Cu, $^{144}$Ce--$^{144}$Pr and $^{20}$F in the mass range of interest provide slightly weaker upper bounds (between $10^{-3}$ and $10^{-2}$) compared to $^{63}$Ni and $^{35}$S at lower masses.

\subsection{Sterile Neutrino Decays}

A sterile neutrino produced in the beta decay of a neutron-rich isotope in a reactor or a light element in the sun can decay before detection via the channels $N\to \nu\nu\bar{\nu}$ and $N\to e^{+}e^{-}\nu$. The former channel is mediated by a neutral current and the latter via either a neutral or charged current. The latter also requires the sterile neutrino mass to be $m_N > 2m_e$. At tree-level (in the single-generation case) the total decay rate is given approximately by
\begin{align}
	\Gamma^{\text{tot}}\approx 2\times\frac{G_F^2}{96\pi^3}|V_{eN}|^2 m_N^5\,,
\end{align}
where the factor of 2 is present in the Majorana case. For RH currents, the factor $|V_{eN}|^2$ is replaced by $|\epsilon_{LR}|^2$ or $|\epsilon_{RR}|^2$.

Reactor experiments with neutrino energies $\sim 10$~MeV are sensitive to sterile neutrinos with masses in the range $1~\text{MeV} < m_N < 10~\text{MeV}$. Limits have been set by searches at the Rovno~\cite{Derbin:1993wy} and Bugey~\cite{PhysRevD.52.1343} reactors. Sterile neutrino decays were also searched for by the Borexino experiment~\cite{PhysRevD.88.072010} which was sensitive to heavy neutrinos with masses up to $14$~MeV produced in the decays of solar $^8$B nuclei. Borexino enforces the relatively stringent limit $|V_{eN}|^2\lesssim 10^{-6} - 10^{-5}$ for $m_N\sim 10$~MeV.

\subsection{Cosmological and Astrophysical Constraints}

The presence of sterile states with mixing strengths $|V_{\ell N}|^2$ (and/or the presence of RH currents) has wide-ranging consequences for early-universe observables. These include the abundances of light nuclei formed during Big Bang Nucleosynthesis (BBN), temperature anisotropies in the Cosmic Microwave Background (CMB) radiation and the large-scale clustering of galaxies ~\cite{Abazajian:2012ys}. Deviations from the standard smooth, isotropic background evolution (and perturbations around this background) impose severe constraints -- the region between the grey lines labelled $\text{CMB}+\text{BAO}+H_0$ (an upper limit) and $\mathrm{BBN}$ (a lower limit) is excluded. These limits are highly sensitive however to the production and decay mechanism of the sterile state and can be relaxed in certain models. 

The main constraint to consider in the $0.1~\text{MeV} < m_N < 3$~MeV mass range is the upper limit labelled $\text{CMB}+\text{BAO}+H_0$. Via the active-sterile mixing or RH current, sterile states are populated in the early-universe and they decouple when the Hubble expansion overcomes the interaction rate with the SM particles. It is then possible for these states to decay at later times to produce non-thermally distributed active neutrinos, modifying the amount of extra radiation measured at recombination, $\Delta N_{\mathrm{eff}}$, beyond the usual value including active neutrino oscillations, $N_{\text{eff}} \simeq 3.046$. Useful probes include the CMB shift parameter $R_{\mathrm{CMB}}$, the first peak of the Baryon Acoustic Oscillations (BAO) and the Hubble parameter $H(z)$ inferred from type Ia supernovae, BAO and Lyman-$\alpha$ data. These exclude values of $m_N$ and $|V_{eN}|^2$ corresponding to lifetimes up to the present day, where the condition that $N$ does not make up more than the observed matter density $\Omega_{\text{sterile}} < \Omega_{\text{DM}} \approx 0.12\,h^{-2}$ also applies. This constraint can be evaded in exotic models~\cite{Bezrukov:2009th, Nemevsek:2012cd, El-Zant:2013nta, Biswas:2018iny}, for example those that inject additional entropy and dilute the dark matter (DM) energy density.

\section{Double Beta Decay Rate with a Sterile Neutrino}
\label{sec:dbd}

Considering one sterile neutrino $N$ with mass $m_N < Q_{\beta\beta} \lesssim$~few~MeV and a SM charged-current as in Eq.~\eqref{eq:lagrangian} with additional suppression by the active-sterile mixing strength $V_{eN}$ allows for the possibility that in $2\nu\beta\beta$ decay one $\bar N$ is emitted ($\nu N\beta\beta$) instead of a $\bar\nu_e$ (we assume that $N$ is long-lived and does not decay within the detector, thus being invisible). The final state is different from the standard $2\nu\beta\beta$ decay and thus there is no interference between $\nu N\beta\beta$ and $2\nu\beta\beta$. There is also no anti-symmetrisation with respect to the two different neutrinos in $\nu N\beta\beta$. Moreover, a RH lepton current can be also assumed to be associated with the emission of the sterile neutrino, which further affects the $\vvbb$ observables, mainly the angular correlation of the outgoing electrons. 

In order to write down expressions for the $\vvbb$ and $\nu N\beta\beta$ decay rates, including the possibility of RH currents, let us start with the general expression \cite{Doi:1985dx}
\begin{align}
\label{eq:totdiffrate}
	d\Gamma =
	2(2-\delta_{\bar\nu_i\bar\nu_j})\pi\delta(E_{e_1}+E_{e_2}+E_{\bar\nu_1}+E_{\bar\nu_2}+E_f-E_i)
	\sum_\text{spins} |\mathcal{R}^{2\nu}|^2
	d\Omega_{e_1} d\Omega_{e_2} d\Omega_{\bar\nu_1} d\Omega_{\bar\nu_2},
\end{align}
where $E_i$, $E_f$, $E_{e_i} = \sqrt{p_{e_i}^2 + m^2_e}$ and $E_{\bar\nu_i} = \sqrt{p_{\bar\nu_i}^2 + m^2_{\nu_i}}$ ($i = 1, 2$) denote the energies of initial and final nuclei, electrons and antineutrinos, respectively. The magnitudes of the associated spatial momenta are $p_{e_i} = |\vecl{p}_{e_i}|$ and $p_{\bar\nu_i} = |\vecl{p}_{\bar\nu_i}|$ and $m_e$ and $m_{\nu_i}$ denote the electron and neutrino masses. The phase space differentials are $d\Omega_{e_1} = {d^3\mathbf{p}_{e_1}}/(2\pi)^3$, etc.. The symmetry factor in Eq.~\eqref{eq:totdiffrate} is $(2-\delta_{\bar\nu_i\bar\nu_j}) = 1$ if identical neutrinos are being emitted in the process and $(2-\delta_{\bar\nu_i\bar\nu_j}) = 2$ if they are distinguishable, i.e. in the case of $\nu N\beta\beta$. Here, the amplitude $\mathcal{R}^{2\nu}$ contains the average contribution from two diagrams with the neutrinos interchanged, with a relative minus sign if the neutrinos are identical. Note that in our calculations we neglect the mass of the light neutrino being emitted and we retain only the mass $m_N$ of the heavy neutrino.

After integrating over the phase space of the outgoing neutrinos, the resulting differential $\vvbb$ decay rate can be generally written in terms of the energies $0 \leq E_{e_1}$, $E_{e_2} \leq Q + m_e$ of the two outgoing electrons, with $Q = E_i - E_f - 2m_e$, and the angle $0 \leq \theta \leq \pi$ between the electron momenta $\vecl{p}_{e_1}$ and $\vecl{p}_{e_2}$
as \cite{Doi:1985dx}
\begin{align}
\label{eq:diffrate}
	\frac{d\Gamma^{2\nu}}{dE_{e_1} dE_{e_2} d\!\cos\theta} 
	= \frac{c_{2\nu}}{2} \left(A^{2\nu} 
	+ B^{2\nu} \cos\theta \right) 
	p_{e_1} E_{e_1} p_{e_2} E_{e_2},
\end{align}
where
\begin{align}
\label{eq:constant}
	c_{2\nu} = (2-\delta_{\bar\nu_i\bar\nu_j})\frac{G_{\beta}^4m_e^9}{8\pi^7},
\end{align}
with $G_\beta = G_F\cos{\theta_C}$ ($G_F$ is the Fermi constant and $\theta_C$ is the Cabbibo angle).

The quantities $A^{2\nu}$ and $B^{2\nu}$ in Eq.~\eqref{eq:diffrate}, generally functions of the electron energies, include the integration over the neutrino phase space,
\begin{align}
\label{eq:ABnuint}
	A^{2\nu} &= 
	\int_{m_{\nu_1}}^{E_i-E_f-E_{e_1}-E_{e_2}} \mathcal{A}^{2\nu}~ \sqrt{E_{\bar\nu_1}^2-m_{\bar\nu_1}^2} \sqrt{(E_i - E_f - E_{e_1} - E_{e_2} - E_{\bar\nu_1})^2 - m_{\nu_2}^2} \nn \\
        &\phantom{=} \times E_{\bar\nu_1} (E_i - E_f - E_{e_1} - E_{e_2} - E_{\bar\nu_1})~ dE_{\bar\nu_1}\,, \\ 
	B^{2\nu} &= 
	\int_{m_{\nu_1}}^{E_i-E_f-E_{e_1}-E_{e_2}} \mathcal{B}^{2\nu}~
        \sqrt{E_{\bar\nu_1}^2-m_{\bar\nu_1}^2} \sqrt{(E_i - E_f - E_{e_1} - E_{e_2} - E_{\bar\nu_1})^2 - m_{\nu_2}^2} \nn \\
        &\phantom{=} \times E_{\bar\nu_1} (E_i - E_f - E_{e_1} - E_{e_2} - E_{\bar\nu_1})~ dE_{\bar\nu_1}\,,
\end{align}
where we have used $E_{\bar\nu_2} = E_i - E_f - E_{e_1} - E_{e_2} - E_{\bar\nu_1}$ due to energy conservation and kept the dependence on the neutrino masses, although in the SM case they can be safely neglected. In turn, the quantities ${\cal A}^{2\nu}$ and ${\cal B}^{2\nu}$, generally functions of the electron and neutrino energies, are calculated below using the nuclear and leptonic matrix elements. 

The rate corresponding to $\nu N\beta\beta$ decay then differs only by the non-negligible mass of the sterile neutrino entering the neutrino energy and, most importantly, the integration bounds. Consequently, the corresponding rate can be obtained from the above by a simple substitution $\nu_1 \to N$, $\nu_2 \to \nu$ and neglecting the mass $m_\nu$. As shown later in this section, in the standard case with only LH lepton currents the quantities ${\cal A}^{2\nu}$ and ${\cal B}^{2\nu}$ do not depend on neutrino masses; hence, the main effect of the sterile neutrino mass is the shrunk electron energy distribution given by the effectively smaller $Q$ value, now given by $Q = E_i - E_f - 2m_e - m_N$.

In our calculations we take the $S_{1/2}$ spherical wave approximation for the outgoing electrons, i.e.
\begin{align}
\label{eq:radialwf}
	\psi_s({p}_e) = \begin{pmatrix} 
		g_{-1}(E_e)\chi_s \\ 
		 f_{+1}(E_e)\left(\vecs{\sigma}\cdot\hat{\vecl{p}}_e\right)\chi_s 
	\end{pmatrix}.
\end{align}
Here, $\hat{\vecl{p}}_e=\vecl{p}_e/|\vecl{p}_e|$ denotes the direction of the electron momentum, $\chi_s$ is a two-component spinor and $g_{-1}(E_e)$ and $f_{+1}(E_e)$ stand for the radial electron wave functions depending on the electron energy $E_e$. As commonly done, we approximate them with their values at the nucleus' surface, i.e. at distance $R$ from the centre of the nucleus. The neutrinos, being neutral, can be simply described as plane waves in the long-wave approximation,
\begin{align}
  \psi(p_\nu) =
  \sqrt{\frac{{E_\nu + m_\nu}}{{2 E_\nu}}}
  \left(\begin{array}{c} 
		      \chi_s \\ 
		      \frac{\left(\vecs{\sigma}\cdot\hat{\mathbf{p}}_\nu\right)}{E_\nu + m_\nu} \chi_s
	\end{array}\right).
\label{eq:vplane} 
\end{align}

\subsection{Purely Left-Handed Currents}
The standard contribution to $\vvbb$ decay given by the first term in the Lagrangian in Eq.~\eqref{eq:lagrangian} has been studied in great detail~\cite{Haxton:1985am,Simkovic:2018rdz}. Sticking to the formalism outlined above, the decay rate is described by the functions
\begin{align}
\label{eq:A-SM}
	{\cal A}^{2\nu}_{\rm SM} &=
	\bigg\{\frac{1}{4}\left[g_V^2 \left(M^K_{F} + M^L_{F}\right)
	- g_A^2 \left(M^K_{GT} + M^L_{GT}\right)  \right]^2 \nn\\
	&~~\, + \frac{3}{4}\left[ g_V^2 \left(M^K_{F} - M^L_{F}\right)
	+ \frac{1}{3} g_A^2 \left(M^K_{GT} - M^L_{GT}\right) \right]^2 \bigg\} \nn\\
	&~~\, \times [g_{-1}^2(E_{e_1})+f_1^2(E_{e_1})][g_{-1}^2(E_{e_2})+f_1^2(E_{e_2})]\,,
\end{align}
and
\begin{align}
\label{eq:B-SM}
	{\cal B}^{2\nu}_{\rm SM} &=
	\bigg\{ \frac{1}{4}\left[ g_V^2 \left(M^K_{F} + M^L_{F}\right)
	- g_A^2 \left(M^K_{GT} + M^L_{GT}\right)  \right]^2 \nn \\
	&~~\, - \frac{1}{4}\left[ g_V^2 \left(M^K_{F} - M^L_{F}\right) + \frac{1}{3} g_A^4 \left(M^K_{GT} - M^L_{GT}\right) \right]^2 \bigg\} \nn \\
	&~~\, \times 4f_1(E_{e_1})f_1(E_{e_2})g_{-1}(E_{e_1})g_{-1}(E_{e_2})\,,
\end{align}
where we define Fermi and Gamow-Teller nuclear matrix elements
\begin{align}
\label{fagtmatt}
	M^{K,L}_{F, GT} = m_e \sum_n M_{F, GT}(n)
	\frac{E_n - (E_i+E_f)/2}{[E_n - (E_i+E_f)/2]^2 - \varepsilon^2_{K,L}}\,.
\end{align}
The electron mass $m_e$ in the above expression is inserted conventionally to make the nuclear matrix elements dimensionless.
The lepton energies enter in Eq.~\eqref{fagtmatt} through the terms
\begin{align}
\label{eq:e-denominators}
  \varepsilon_K = 
  \frac{1}{2}\left(E_{e_2}+E_{\bar\nu_2}-E_{e_1}-E_{\bar\nu_1}\right)\,,\qquad
  \varepsilon_L = 
  \frac{1}{2}\left(E_{e_1}+E_{\bar\nu_2}-E_{e_2}-E_{\bar\nu_1}\right)\,,
\end{align}  
which satisfy $-Q/2 \leq \varepsilon_{K,L} \leq Q/2$. In case of $2\nu\beta\beta$ decay with energetically forbidden transitions to the intermediate states, $E_n - E_i > - m_e$, the quantity $E_n -(E_i+E_f)/2 = Q/2 + m_e + (E_n-E_i)$ is always larger than $Q/2$.

The above expressions may be further simplified using several well-motivated approximations.

\paragraph*{Isospin Invariance:} Neglecting the isospin non-conservation in the nucleus, the double Fermi nuclear matrix elements vanish, i.e. $M_F^K = M_F^L = 0$. Therefore, Eqs.~\eqref{eq:A-SM} and \eqref{eq:B-SM} then respectively acquire the approximate form 
\begin{align}
\label{eq:A-SM}
	{\cal A}^{2\nu}_{\rm SM} &\approx
	\frac{1}{4}g_A^4\left[ \left(M^K_{GT} + M^L_{GT}\right)^2 + \frac{1}{3}  \left(M^K_{GT} - M^L_{GT}\right)^2 \right] \nn\\
	&~~\, \times [g_{-1}^2(E_{e_1})+f_1^2(E_{e_1})][g_{-1}^2(E_{e_2})+f_1^2(E_{e_2})]\,,
\end{align}
and
\begin{align}
\label{eq:B-SM}
	{\cal B}^{2\nu}_{\rm SM} &\approx
	\frac{1}{4}g_A^4\left[ \left(M^K_{GT} + M^L_{GT}\right)^2 + \frac{1}{9} \left(M^K_{GT} - M^L_{GT}\right)^2 \right] \nn \\
	&~~\, \times 4f_1(E_{e_1})f_1(E_{e_2})g_{-1}(E_{e_1})g_{-1}(E_{e_2})\,.
\end{align}

\paragraph*{Nuclear matrix element dependence on lepton energies:} If we neglect the dependence of nuclear matrix elements on $\varepsilon_{K,L}$, the nuclear and leptonic parts can be separated and we get
\begin{align}\label{nucmata}
	{\cal A}^{2\nu}_{\rm SM} &\approx
	g_A^4 M_{GT}^2 [g_{-1}^2(E_{e_1})+f_1^2(E_{e_1})][g_{-1}^2(E_{e_2})+f_1^2(E_{e_2})]\,, \\
	{\cal B}^{2\nu}_{\rm SM} &\approx
	g_A^4 M_{GT}^2 4f_1(E_{e_1})f_1(E_{e_2})g_{-1}(E_{e_1})g_{-1}(E_{e_2})\,,
\end{align}
with the Gamow-Teller nuclear matrix element now defined as
\begin{align}
	M_{GT} &= m_e \sum_n 
	\frac{\langle 0^+_f | \sum_{m}\tau^+_m 
		\sigma_m | 1^+_{n}\rangle
	\langle 1^+_n | \sum_{m}\tau^+_m \sigma_m | 0^+_{i} \rangle}
	{E_n - {(E_i+E_f)}/{2}}\,.
\end{align}

A better approximation is obtained by Taylor expansion of the nuclear matrix elements in the small parameters $\epsilon_{K,L}$~\cite{Simkovic:2018rdz}. Keeping terms up to the fourth power in $\epsilon_{K,L}$ gives
\begin{align}
\label{eq:expampA}
	{\cal A}^{2\nu}_{\text{SM}} &\approx
	g_A^4 \bigg[ (M_{GT-1})^2 + (\epsilon_K^2+\epsilon_L^2)M_{GT-1}M_{GT-3} + \frac{1}{3}\epsilon_K^2 \epsilon_L^2 (M_{GT-3})^2 \nn \\
	&\hspace{.65cm} + (\epsilon_K^4+\epsilon_L^4)\left(M_{GT-1}M_{GT-5}+\frac{1}{3}(M_{GT-3})^2\right) \bigg] \nn \\
		&\hspace{.65cm} \times [g_{-1}^2(E_{e_1})+f_1^2(E_{e_1})][g_{-1}^2(E_{e_2})+f_1^2(E_{e_2})]\,,
\end{align}
and
\begin{align}
\label{eq:expampB}
	{\cal B}^{2\nu}_{\text{SM}} &\approx
	g_A^4 \bigg[ (M_{GT-1})^2 + (\epsilon_K^2+\epsilon_L^2)M_{GT-1}M_{GT-3} + \frac{4}{9}\epsilon_K^2 \epsilon_L^2 (M_{GT-3})^2 \nn \\
	&\hspace{.65cm} + (\epsilon_K^4+\epsilon_L^4)\left(M_{GT-1}M_{GT-5}+\frac{5}{18}(M_{GT-3})^2\right) \bigg] \nn \\
	&\hspace{0.65cm}\times 4f_1(E_{e_1})f_1(E_{e_2})g_{-1}(E_{e_1})g_{-1}(E_{e_2})\,.
\end{align}
Here, the nuclear matrix elements introduced are defined as 
\begin{align}
	M_{GT-1} &= M_{GT}\,, \\
	M_{GT-3} &= m_e^3 \sum_n \frac{4M_{GT}(n)}{(E_n-(E_i+E_f)/2)^3}\,, \\
	M_{GT-5} &= m_e^5 \sum_n \frac{16M_{GT}(n)}{(E_n-(E_i+E_f)/2)^5}\,.
\end{align}
This is the approximation we employ in our later numerical analyses.

\subsection{Contribution with a Right-Handed Current}
The non-standard contribution to $\vvbb$ decay involving the RH currents proportional to the $\epsilon_{XR}$ coupling, as appearing in the Lagrangian in Eq.~\eqref{eq:lagrangian}, was calculated in Ref.~\cite{Deppisch:2020mxv}. The corresponding functions $A^{2\nu}$ and $B^{2\nu}$ entering Eq.~\eqref{eq:diffrate} read
\begin{align}
	{\cal A}^{2\nu}_{\epsilon} &=
  	4 \bigg\{ \left[ g_V^4 (M_F^K - M_F^L)^2 
  	+ \frac{1}{3}g_A^4 (M_{GT}^K - M_{GT}^L)^2 \right] \nn\\
  	&\quad\,\,\,+\left[g_V^4 (M_F^K + M_F^L)^2 + \frac{1}{3}g_A^4 (M_{GT}^K 
  	 + M_{GT}^L)^2 \right] \bigg\} \nn\\
	&\quad\times \bigg\{
	 [g_{-1}^2(E_{e_1})+f_1^2(E_{e_1})][g_{-1}^2(E_{e_2})+f_1^2(E_{e_2})] \nn\\
	&\quad\,\,\,+[g_{-1}^2(E_{e_1})-f_1^2(E_{e_1})]
	  [g_{-1}^2(E_{e_2})-f_1^2(E_{e_2})]
	  \frac{m_{\nu}m_{N}}{E_{\nu_1} E_{\nu_2}} \bigg\} \nn\\
	&+2\bigg\{ \left[ g_V^4 (M_F^K - M_F^L)^2 
	 -\frac{1}{3}g_A^4 (M_{GT}^K - M_{GT}^L)^2 \right] \nn\\
	&\quad\,\,\,-\left[ g_V^4 (M_F^K + M_F^L)^2 - \frac{1}{3}g_A^4 (M_{GT}^K 
	 + M_{GT}^L)^2 \right] \nn\\ 
	&\quad\,\,\,+ 2 g_V^2g_A^2 \left[(M_F^K - M_F^L)(M_{GT}^K - M_{GT}^L) 
	 + (M_F^K + M_F^L)(M_{GT}^K + M_{GT}^L) \right] \bigg\} \nn\\
	&\quad\times \bigg\{ 
	 [g_{-1}^2(E_{e_1})-f_1^2(E_{e_1})][g_{-1}^2(E_{e_2})-f_1^2(E_{e_2})] \nn\\
	&\quad\,\,\,+[g_{-1}^2(E_{e_1})+f_1^2(E_{e_1})]
	  [g_{-1}^2(E_{e_2})+f_1^2(E_{e_2})]\frac{m_{\nu}m_{N}}{E_{\nu_1} E_{\nu_2}} \bigg\}\,.
\label{eq:Aright}
\end{align}
Here, the dependence on the electron radial wave functions has been made explicit. Likewise, the terms proportional to $\hat{\vecl{p}}_1\cdot \hat{\vecl{p}}_2 = \cos\theta$ combine to give
\begin{align}
	{\cal B}^{2\nu}_{\epsilon} &=
	\bigg\{ 2 g_V^4 \left[ (M_F^K + M_F^L)^2 - (M_{F}^K - M_{F}^L)^2 \right]
	\frac{m_{\nu}m_{N}}{E_{\nu_1} E_{\nu_2}} \nn\\
	&\quad+\frac{8}{9} g_A^4 \left[ (M_{GT}^K - M_{GT}^L)^2 + (M_{GT}^K 
	 + M_{GT}^L)^2 \right]  \nn\\
	&\quad+ \frac{10}{9} g_A^4 \left[ (M_{GT}^K + M_{GT}^L)^2 
	 - (M_{GT}^K - M_{GT}^L)^2 \right] \frac{m_{\nu}m_{N}}{E_{\nu_1} E_{\nu_2}} \nn\\
	&\quad+ \frac{4}{3} g_V^2g_A^2 \left[(M_F^K - M_F^L)(M_{GT}^K 
	 - M_{GT}^L)
	 + (M_F^K + M_F^L)(M_{GT}^K + M_{GT}^L) \right]
	 \frac{m_{\nu}m_{N}}{E_{\nu_1} E_{\nu_2}} \nn \\
	&\quad- \frac{8}{3} g_V^2g_A^2 \left[(M_F^K - M_F^L)(M_{GT}^K 
	 - M_{GT}^L) + (M_F^K + M_F^L)(M_{GT}^K + M_{GT}^L) \right]
	 \bigg\} \nn\\
	 &\times 4f_1(E_{e_1})f_1(E_{e_2})g_{-1}(E_{e_1})g_{-1}(E_{e_2})\,,
\label{eq:Bright}
\end{align}
In Eqs.~\eqref{eq:Aright} and \eqref{eq:Bright}, the terms proportional to $m_{\nu}m_{N}$ are small, as one of the emitted neutrinos is still assumed to be the light with $m_\nu \lesssim 0.1$~eV. As in the SM case, for the purpose of numerical computations we approximate the above expressions with their Taylor expansions up to the fourth power in the small parameters $\epsilon_{K,L}$.

\subsection{Decay Distributions and Total Rate}
The kinematics of the electrons emitted in the decay is captured by the fully differential decay rate expressed in Eq.~\eqref{eq:diffrate} depending on the (in principle) observable electron energies $E_{e_1}$, $E_{e_2}$ and the angle $\theta$ between the electron momenta. All the information is contained by the quantities $\mathcal{A}^{2\nu}$ and $\mathcal{B}^{2\nu}$ presented above both for the standard LH (Eqs.~\eqref{eq:A-SM} and \eqref{eq:B-SM}) and the exotic RH (Eqs.~\eqref{eq:expampA} and \eqref{eq:expampB}) case. The following values of the physical constant are used in our numerical computations: $G_\beta = 1.1363 \times 10^{-11}$~MeV$^{-2}$, $\alpha = 1/137$, $m_e = 0.511$~MeV, $m_p = 938$~MeV, $R = 1.2A^{1/3}$~fm (nucleon number $A=100$ for Molybdenum), $Q({}^{100}\mathrm{Mo})=3.03$~MeV, $g_V = 1$. Since quenching of the axial coupling $g_A$ is expected in the nucleus~\cite{Gysbers:2019uyb}, we take $g_A = 1$ instead of the usual value $g_A^{\rm nucleon}=1.269$ for a free neutron. Further, we use the $\vvbb$ decay nuclear matrix elements from Ref.~\cite{Simkovic:2018rdz}, as shown in Tab.~\ref{tab:nmes}.

\begin{table}[!t]
\centering 
\renewcommand{\arraystretch}{1.1}
\setlength\tabcolsep{0.2cm}
\begin{tabular}{rlllcccc}\hline\hline
Isotope & $M^{2\nu}_{GT-1}$ & $M^{2\nu}_{GT-3}$ & $M^{2\nu}_{GT-5}$ \\ \hline
$^{76}$Ge  & $0.111$  & $0.0133$  & $0.00263$ \\
$^{82}$Se  & $0.0795$ & $0.0129$  & $0.00355$ \\
$^{100}$Mo & $0.184$
& $0.0876$
& $0.0322$ 
\\
$^{136}$Xe & $0.0170$ & $0.00526$ & $0.00169$ \\
\hline\hline
\end{tabular}
\caption{Nuclear matrix elements calculated within the pn-QRPA with partial isospin restoration \cite{Simkovic:2018rdz} assuming the effective axial coupling $g_A = 1.0$.
\label{tab:nmes}}
\end{table}

With all the above ingredients we can now calculate the the total decay rate as well as various decay distributions potentially observable in $2\nu\beta\beta$ decay experiments.

\paragraph*{Total electron energy and single electron energy:} 

The $2\nu\beta\beta$ decay experiments measure primarily the distribution with respect to the total kinetic energy of the outgoing electrons, i.e.\ $d\Gamma^{2\nu}/dE_K$ with $E_K = E_{e_1} + E_{e_2} - 2m_e - m_{\nu_1} - m_{\nu_2}$. Here, $m_{\nu_{1,2}}$ denote the masses of the emitted neutrinos, which can be safely neglected in the SM case, but we consider also a contribution involving a heavy sterile neutrino, in which case one of the masses becomes non-negligible and we denote it $m_N$. We also neglect the recoil of the final state isotope which would change the endpoint by $\sim Q^2/M \lesssim 0.1$~keV, with the $Q \lesssim 3$~MeV and the mass of the nucleus $M \approx 76 - 136$~GeV. Some experiments capture the energies and tracks of individual electrons, thus allowing for study of the single electron energy distribution $d\Gamma^{2\nu}/dE_{e_1}$ (the symmetry of the process ensures the distribution with respect to the second electron is identical) and the double differential distribution $d\Gamma^{2\nu}/(dE_{e_1}dE_{e_2})$. These distributions are calculated from Eq.~\eqref{eq:diffrate} as
\begin{align}
	\frac{d\Gamma^{2\nu}}{dE_{e_1}dE_{e_2}} &= \int_{-1}^1 d\cos\theta \,\frac{d\Gamma^{2\nu}}{dE_1 dE_2 d\cos\theta}\,, \nn\\
	\frac{d\Gamma^{2\nu}}{dE_{e_1}} &= \int_{m_e}^{E_i - E_f - m_{\nu_1} - m_{\nu_2} - E_{e_1}}dE_{e_2}
	\,\frac{d\Gamma^{2\nu}}{dE_{e_1}dE_{e_2}}\,, \nn\\
  \frac{d \Gamma^{2\nu}}{d E_K} &= \frac{E_K}{E_K^{\rm max}} \int_{0}^{E_K^{\rm max}} d E~
  \frac{d \Gamma^{2\nu}}{dE_{e_1} dE_{e_2}},
\end{align}
where in the latter
\begin{align}
\label{eq:E-EK}
	E_{e_1} = E_K - \frac{E_K }{E_K^{\rm max}} E + m_e, \quad 
	E_{e_2} = \frac{E_K }{E_K^{\rm max}} E + m_e,
\end{align} 
and $E_K^{\rm max} = E_i - E_f - 2 m_e - m_{\nu_1} - m_{\nu_2}$. We neglect the light neutrino masses $m_{\nu_1} = m_{\nu_2} = 0$ in the SM case and retain only the heavy neutrino mass in the sterile contribution, $m_{\nu_1} = m_N, m_{\nu_2} = 0$. Given the fact that most experiments provide only the $\vvbb$ decay distribution in dependence on the total kinetic energy of the electrons, in the following analysis we focus primarily on this observable.

The kinematic endpoint of the summed electron energy spectrum of $\nu N\beta\beta$ decay with an emission of sterile neutrino is of primary interest, as it leads to a distortion in the spectrum as the main experimental signal. Here we note that the quantity ${\cal A}^{2\nu}$ in Eq.~\eqref{eq:expampA} depends only weakly on the heavy neutrino mass $m_N$ as in the Taylor expansion in the parameters $\epsilon_{K,L}$ the leading term, which is free of $m_N$ and the lepton energies, is the dominant one. By restricting our consideration only to this leading term for the sterile neutrino with left-handed current we can express the energy spectrum as
\begin{align}
	\frac{d\Gamma^{2\nu}}{dE_K} = \frac{E_K}{E^\text{max}_K} f(E_K) F_N(E_K, m_N),
\end{align}
with
\begin{align}
	f(E_K) = c_{2\nu} g_A^4 M^2_{GT-1} 
	\int_0^{E^\text{max}_K} \!\!\!\!\! p_{e_1} E_{e_1} p_{e_2} E_{e_2}
	\left(g_{-1}^2(E_{e_1}) + f^2_1(E_{e_1})\right)
	\left(g_{-1}^2(E_{e_2}) + f^2_1(E_{e_2})\right)dE,
\end{align}
where $E_{e_1}$ and $E_{e_2}$ are expressed in terms of $E_K$ and $E$ according to Eq.~\eqref{eq:E-EK}. The shape of the distribution near the endpoint is determined by the function
\begin{align}
	F_N(E_K,m_N) &=
	\frac{1}{60}\sqrt{(y + m_N)^2 - m_N^2}
	\left[2(y + m_N)^4 - 9(y + m_N)^2 m^2_N - 8m_N^4\right] \nonumber\\
	&+\frac{1}{4}(y + m_N)m^4_N\,
	\ln{\left|\frac{y}{m_N} + 1 + \sqrt{\left(\frac{y}{m_N} + 1\right)^2 - 1}\right|},
\end{align}
with $y = E_K^\text{max} - E_K$ ($0 < y < E_K^\text{max} = Q - m_N$). For $m_N =0$, this function reduces to $F_N(E_K, 0) = (E^\text{max}_K - E_K)^5/30$, leading to the well known scaling of standard $2\nu\beta\beta$ decay near the endpoint.

\begin{figure}[t!]
	\centering
	\includegraphics[width=0.55\textwidth]{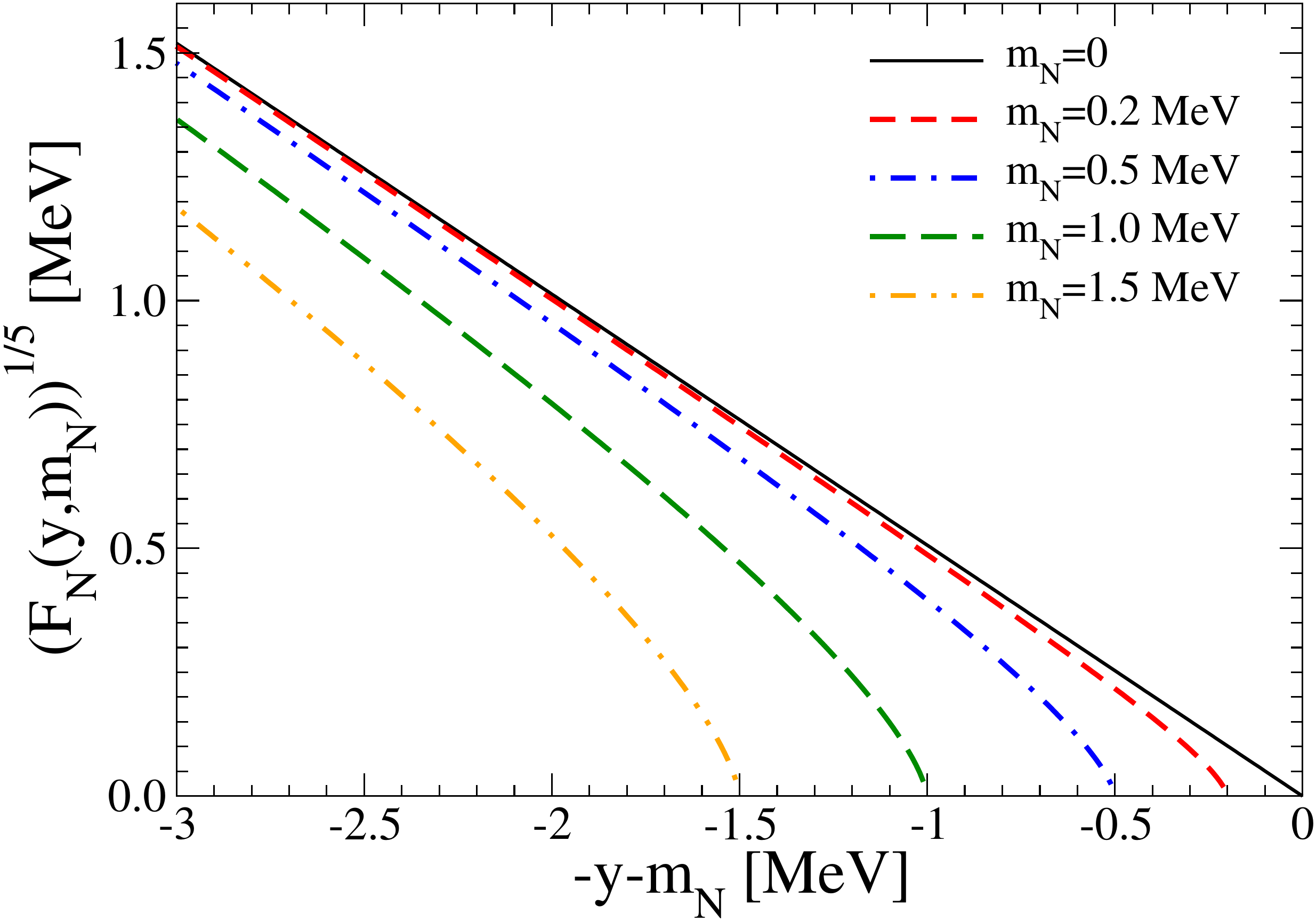}
	\caption{Kurie-type expression ${\cal K}(E_K, m_N) = F_N^{1/5}(E_K, m_N)$ for $\nu N\beta\beta$ decay as a function of $-y-m_N$ for various values of the neutrino mass, $m_N = 0, 0.2, 0.5, 1.0, 1.5$~MeV.}
	\label{fig:kurieF}
\end{figure}
In analogy to the construction of the Kurie function in single $\beta$ decay we introduce the $\nu N\beta\beta$ decay equivalent
\begin{align}
	{\cal K}(E_K, m_N) 
	= \left(\frac{d\Gamma^{2\nu}/dE_K}{f(E_K)}\frac{E_K^\text{max}}{E_K}\right)^{1/5}
	= \left(F_N(E_K,m_N)\right)^{1/5}, 
\end{align}
which is plotted in Fig.~\ref{fig:kurieF} as a function of $-y-m_N$ near the endpoint for various neutrino masses. We see that ${\cal K}(E_K)$ is linear near the endpoint for zero neutrino mass ($m_N = 0$). However, the linearity of the Kurie plot is lost if the sterile neutrino has a non-zero mass with the deviation from the straight line depending on the magnitude of $m_N$.

Near the kinematic endpoint $E_K \lesssim E_K^\text{max} = Q - m_N$, the function $F_N(E_K, m_N)$ asymptotically approaches
\begin{align}
	\frac{d\Gamma^{2\nu}}{dE_K} 
	\propto F_N(E_K, m_N) 
	\xrightarrow[0 < y \ll m_N]{} \frac{16\sqrt{2}}{105}m_N^{3/2}\left(E_K^\text{max} - E_K\right)^{7/2}.
\end{align}
Hence, the total electron energy spectrum of $\nu N\beta\beta$ is rather smooth near the endpoint, unlike in the case of single $\beta$ decay. Therefore, no sharp kink is expected to appear in the total energy spectrum including both the SM and the sterile neutrino contributions.

\paragraph*{Angular correlation factor and total decay rate:}
The integration over the electron energies leads to the equation
\begin{align}
    \label{eq:angular_dist}
	\frac{d\Gamma^{2\nu}}{d\cos\theta} = 
	\frac{\Gamma^{2\nu}}{2}\left(1 + K^{2\nu} \cos\theta\right)\,,
\end{align}
describing the angular distribution of the decay. Here, $\Gamma^{2\nu}$ denotes the total $\vvbb$ decay rate and $K^{2\nu} = \Lambda^{2\nu}/\Gamma^{2\nu}$ stands for the angular correlation factor, which are given by
\begin{align}
 \label{eq:gammaLambda}
	\begin{pmatrix}
    	\Gamma^{2\nu} \\
    	\Lambda^{2\nu}
	\end{pmatrix}  
 	&= \frac{c_{2\nu}}{m_e^{11}}
 	        \int_{m_e}^{E_i-E_f-m_e} dE_{e_1} p_{e_1} E_{e_1}
 	        \int_{m_e}^{E_i-E_f-E_{e_1}} dE_{e_2} p_{e_2} E_{e_2}
    \begin{pmatrix}
    	A^{2\nu} \\
    	B^{2\nu}
  	\end{pmatrix}.
\end{align}
As the inclusion of RH current leads to the opposite sign of the angular correlation of the emitted electrons~\cite{Deppisch:2020mxv}, it can be also used to distinguish the corresponding contributions, as analysed in the following section.

\section{Constraints on Sterile Neutrino Parameters}
\label{sec:results}

We will now use the differential $2\nu\beta\beta$ decay rates derived in Sec. \ref{sec:dbd} to exclude regions of the sterile neutrino parameter space -- namely, the sterile neutrino mass $m_{N}$ and mixing with the electron neutrino $|V_{eN}|^2$. To do this we will first outline a simple frequentist limit setting method. We will then use the non-observation of deviations from the SM $2\nu\beta\beta$ decay spectrum by $0\nu\beta\beta$ decay search experiments such as GERDA-II, CUPID-0, NEMO-3 and KamLAND-Zen to put upper limits on $|V_{eN}|^2$ as a function of $m_{N}$. We will also estimate upper limits from the forecasted sensitivities of future $0\nu\beta\beta$ decay experiments such as LEGEND, SuperNEMO, CUPID and DARWIN. Finally, we will compare these upper limits to existing constraints in the $0.1 ~\mathrm{MeV} < m_{N} < 3$ MeV range from single beta decay probes ($^{64}$Cu, $^{144}$Ce$-^{144}$Pr and $^{20}$F) and sterile neutrino decays (Borexino) as discussed in Sec. \ref{sec:searches}.

\subsection{Statistical Procedure}

To obtain upper limits on the mixing $|V_{eN}|^2$ we follow the standard frequentist approach of Refs.~\cite{Tanabashi:2018oca, Kahlhoefer:2019vhz}. Firstly, we define the total differential $2\nu\beta\beta$ decay rate as the incoherent sum of the sterile neutrino and SM rates for a given sterile mass $m_N$ and total kinetic energy $E_K = E_{e_1} + E_{e_2} - 2m_e$,
\begin{align}
\label{eq:total_dist}
	\frac{d\Gamma^{2\nu}(\boldsymbol{\xi})}{dE_K} 
	= (1 - |V_{eN}|^2)^2 \frac{d\Gamma^{2\nu}_{\text{SM}}}{dE_K} 
	+ (1-|V_{eN}|^2)|V_{eN}|^2 \,\frac{d\Gamma^{2\nu}_N(m_N)}{dE_K},
\end{align}
explicitly writing the dependence on active-sterile mixing $|V_{eN}|^2$. The total differential rate depends on the sterile neutrino parameters $\boldsymbol{\xi} \equiv (m_{N}, |V_{eN}|^2)$ and $E_{K}$. Here, the contribution $d\Gamma_N^{2\nu}/dE_K$ due to the sterile neutrino includes a factor of two compared to the SM contribution, as two distinguishable neutrinos are emitted in the process, cf. Eq.~\eqref{eq:totdiffrate}.

\begin{figure}[t!]
	\centering
	\includegraphics[width=0.9\textwidth]{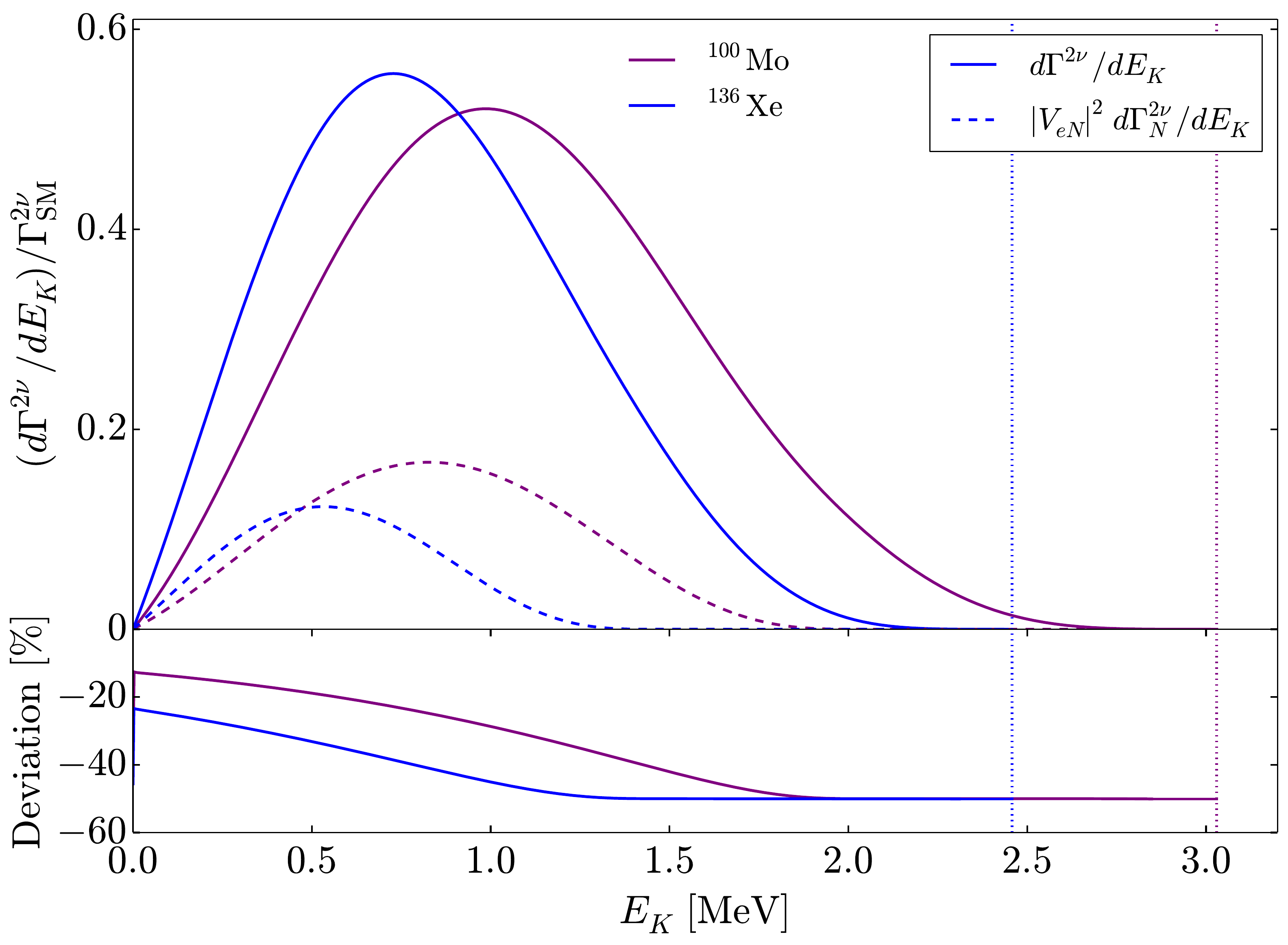}
	\caption{Total differential $2\nu\beta\beta$ decay rate (solid) and the sterile neutrino contribution (dashed) with $m_N = 1.0$ MeV and $|V_{eN}|^2 = 0.5$ for the two isotopes $^{100}$Mo (purple) and $^{136}$Xe (blue). Both distributions are normalised to the SM decay rate. The vertical dotted lines indicate the respective $Q$ values and the panel at the bottom shows the corresponding percentage deviations from the SM rate.}
	\label{fig:isotope_plot}
\end{figure}
In Fig.~\ref{fig:isotope_plot}, the total differential decay rate in Eq.~\eqref{eq:total_dist} is compared to the sterile neutrino contribution $|V_{eN}|^2\cdot d\Gamma^{2\nu}_N/dE_K$ (where both are normalised to the total SM decay rate $\Gamma^{2\nu}_\text{SM}$) for the isotopes $^{100}$Mo and $^{136}$Xe. The respective $Q$ values of the isotopes are indicated by the vertical dotted lines and the values $m_N = 1.0$~MeV and $|V_{eN}|^2 = 0.5$ are chosen. In the panel below we show the corresponding percentage deviation of the total differential rate from the SM rate,
\begin{align}
\label{eq:deviation}
	\left(\frac{d\Gamma^{2\nu}}{dE_K} - \frac{d\Gamma^{2\nu}_\text{SM}}{dE_K}\right)
	\bigg/ \frac{\Gamma^{2\nu}_\text{SM}}{dE_K} 
	= |V_{eN}|^2 \left(\frac{d\Gamma^{2\nu}_N}{dE_K} 
	  \bigg/ \frac{\Gamma^{2\nu}_\text{SM}}{dE_K} - 1\right).
\end{align}
It can be seen that the magnitude of $d\Gamma^{2\nu}/dE_K$ decreases with respect to the $d\Gamma^{2\nu}_\text{SM}/dE_K$ as the total kinetic energy increases, eventually plateauing at around $-10\%$. This is because the sterile neutrino contribution $|V_{eN}|^2\,d\Gamma^{2\nu}_N/dE_K$ falls as $E_K$ increases above $\sim 1.0$ MeV. Eventually its contribution is negligible, but there remains a suppression from the $(1 - |V_{eN}|^2)$ factor multiplying the SM contribution, which is particularly sizeable for the choice $|V_{eN}|^2 = 0.5$. It is apparent from Eq.~\eqref{eq:deviation} that the deviation tends to a factor of $-|V_{eN}|^2$. The characteristic signature of the sterile neutrino is a relative increase of the differential rate for $E_K \lesssim Q - m_N$.

Any experiment measuring the $2\nu\beta\beta$ decay spectrum will count a number of events $N_\text{events}$ distributed over a number of bins $N_\text{bins}$ in the total kinetic energy $E_K$. In the presence of a sterile neutrino, the expected fraction of events $\Delta N^{(i)}_\text{exp}$ per bin will be the integral of $d\Gamma^{2\nu}/dE_K$ over the width of the bin from the total kinetic energy $E_i$ to $E_{i+1}$,
\begin{align}
\label{eq:fraction}
	\Delta N^{(i)}_\text{exp} 
	= \frac{1}{\mathcal{N}} \int_{E_i}^{E_{i+1}} dE_K \,\frac{d\Gamma^{2\nu}}{dE_K}\,, 
\end{align}
where the normalisation factor $\mathcal{N}$ is 
\begin{align}
	\mathcal{N} 
	= \int_{E_\text{min}}^{E_\text{max}} dE_K \,\frac{d\Gamma^{2\nu}}{dE_K}\,,
\end{align}
i.e. the total area enclosed by $d\Gamma^{2\nu}/dE_K$ between kinetic energies $E_\text{min}$ and $E_\text{max}$. The total number of expected events per bin will then be
\begin{align}
	N^{(i)}_\text{exp} 
	= N^{(i)}_\text{sig} + N^{(i)}_\text{bkg} 
	= N_\text{events}\cdot\Delta N^{(i)}_\text{exp}\,,
\end{align}
where we have also split the expected number of events into the number of signal and background events as
\begin{align}
\label{eq:sig_back}
	N^{(i)}_{\text{sig}} 
	&= \frac{N_\text{events}}{\mathcal{N}} |V_{eN}|^2 \int_{E_i}^{E_{i+1}} dE_K 	
	   \left(\frac{d\Gamma_N^{2\nu}}{dE_K} - \frac{d\Gamma_\text{SM}^{2\nu}}{dE_K}\right), \\
	N^{(i)}_\text{bkg} 
	&= \frac{N_\text{events}}{\mathcal{N}}\int_{E_i}^{E_{i+1}} dE_K
	   \,\frac{d\Gamma_\text{SM}^{2\nu}}{dE_K}\,.
\end{align}
The probability of the experiment observing $N^{(i)}_\text{obs}$ events per bin given $N^{(i)}_\text{exp}$ expected events is the Poisson probability $P(N^{(i)}_\text{obs} |N^{(i)}_\text{exp})$. The likelihood of the data $\boldsymbol{\mathrm{D}}$ given the sterile neutrino hypothesis, $\mathcal{L}(\boldsymbol{\mathrm{D}}|\boldsymbol{\xi})$, is defined as the product of the Poisson probabilities over all bins. It is more convenient to write the log-likelihood
\begin{align}
\label{eq:log_likelihood}
	-2\log\mathcal{L}(\boldsymbol{\mathrm{D}}|\boldsymbol{\xi})
	&= 2\sum^{N_\text{bins}}_i
	   \left\{N^{(i)}_\text{exp}(\boldsymbol{\xi}) - N^{(i)}_\text{obs} 
	        + N^{(i)}_\text{obs}
	          \log \left(
	          \frac{N^{(i)}_\text{obs}}{N^{(i)}_\text{exp}(\boldsymbol{\xi})}\right)\right\} 
	\nonumber\\
	&\approx \sum^{N_\text{bins}}_i 
	 \frac{\left(N^{(i)}_\text{obs} - 
	 N^{(i)}_\text{exp}(\boldsymbol{\xi})\right)^2}{N^{(i)}_\text{exp}(\boldsymbol{\xi})}\,,
\end{align}
where the second equality holds via Wilks' theorem if there are a large number of events per bin \cite{Wilks:1938dza}. From this we can construct the test-statistic
\begin{align}
\label{eq:test_statistic}
	q_{\boldsymbol{\xi}}
	&= -2\left(\log \mathcal{L}(\boldsymbol{\mathrm{D}}|\boldsymbol{\xi}) 
	          -\log \mathcal{L}(\boldsymbol{\mathrm{D}}|\hat{\boldsymbol{\xi}})\right),
\end{align}
where $\hat{\boldsymbol{\xi}}$ are the values of the sterile neutrino parameters that minimise the log-likelihood function. The quantity $q_{\boldsymbol{\xi}}$ is expected to follow a $\chi^2$ distribution with one degree of freedom.

We assume that the experiment does not observe a spectrum deviating significantly from the SM prediction. We therefore set the number of observed events in Eq.~\eqref{eq:log_likelihood} to $N^{(i)}_\text{obs} = N^{(i)}_\text{exp}(\boldsymbol{\xi})$ with $\boldsymbol{\xi} = (m_N,0)$. In reality, however, the experiment could be repeated many times and record a different value of $N^{(i)}_\text{obs}$ each iteration. This fluctuation can be imitated by running a series of toy Monte Carlo simulations of the experiment. For every toy Monte Carlo there is a value of $q_{\boldsymbol{\xi}}$, with the relevant test-statistic becoming the median of these values. A representative data set is commonly used as a good approximation of the Monte Carlo method in the large sample limit \cite{Cowan:2010js}. This is the so-called Asimov data set $\boldsymbol{\mathrm{D}}_\text{A}$ for which the observed number of events per bin $N^{(i)}_\text{obs}$ equals the number of background events $N_\text{bkg}^{(i)}$  \cite{Burns:2011xf}. The $\hat{\boldsymbol{\xi}}$ that minimises the log-likelihood to $-2\log \mathcal{L}(\boldsymbol{\mathrm{D}}_\text{A}|\hat{\boldsymbol{\xi}}) = 0$ is then simply $\hat{\boldsymbol{\xi}} = (m_N,0)$ which matches our initial approach.

The magnitude of the test-statistic $q_{\boldsymbol{\xi}} = -2\log\mathcal{L}(\boldsymbol{\mathrm{D}}_\text{A}|\boldsymbol{\xi})$ translates to a degree of compatibility between the Asimov data set and the sterile neutrino hypothesis with parameters $\boldsymbol{\xi} = (m_N, |V_{eN}|^2)$. For example, if both parameters are allowed to vary, combinations of the parameters giving $q_{\boldsymbol{\xi}} \gtrsim 4.61$ are excluded at 90\% confidence level (CL). Rather than performing a two-dimensional scan of the parameters, we instead fix $m_N$ for values over the range $\sim 0.1 - 3$~MeV and find the value of $|V_{eN}|^2$ for which $q_{\boldsymbol{\xi}} = 2.71$, corresponding to the 90\% CL upper limit on the mixing.

Finally we note that we have not yet included the effect of systematic uncertainties. Systematics altering the total number of observed events without leading to distortions in the spectrum can be accounted for by introducing the nuisance parameter $\eta$
\begin{align}
\label{eq:normalisation}
	-2\log\mathcal{L}(\boldsymbol{\mathrm{D}}|\boldsymbol{\xi},\eta)
	\approx \sum^{N_\text{bins}}_i 
	\frac{\left(N^{(i)}_\text{bkg} - 
	(1+\eta)N^{(i)}_\text{exp}(\boldsymbol{\xi})\right)^2}{(\sigma_\text{stat}^{(i)})^2 
	+ (\sigma_\text{sys}^{(i)})^2} + \left(\frac{\eta}{\sigma_\eta}\right)^2,
\end{align}
where $\sigma_\eta$ is a small associated uncertainty. The remaining systematic uncertainties are included in the quantity $\sigma_\text{sys}^{(i)} = \sigma_f N_\text{exp}^{(i)}$ which adds in quadrature with the statistical uncertainty $(\sigma_\text{stat}^{(i)})^2 = N_\text{exp}^{(i)}$ in the denominator of Eq.~\eqref{eq:normalisation}. The test-statistic becomes
\begin{align}
	q_{\boldsymbol{\xi}} 
	= -2\left(\log\mathcal{L}(\boldsymbol{\mathrm{D}}|\boldsymbol{\xi},\hat{\hat{\eta}})
	-\log\mathcal{L}(\boldsymbol{\mathrm{D}}|\hat{\boldsymbol{\xi}},\hat{\eta})\right),
\end{align}
where $\hat{\hat{\eta}}$ minimises the log-likelihood for a given $\boldsymbol{\xi}$ while $\hat{\boldsymbol{\xi}}$ and $\hat{\eta}$ are the values at the global minimum of the log-likelihood. For the Asimov data set the parameters at the global minimum are $\hat{\boldsymbol{\xi}} = (m_N,0)$ and $\hat{\eta} = 0$ such that $-2\log\mathcal{L}(\boldsymbol{\mathrm{D}}_\text{A}|\hat{\boldsymbol{\xi}},\hat{\eta}) = 0$. The test-statistic then reduces to
\begin{align}
\label{eq:finalteststatistic}
	q_{\boldsymbol{\xi}} 
	= \underset{\eta}{\text{min}}\left[\sum^{N_\text{bins}}_i 
	\frac{\left(N^{(i)}_\text{bkg} - (1+\eta)N^{(i)}_\text{exp}(\boldsymbol{\xi})\right)^2}{(\sigma_\text{stat}^{(i)})^2 
	+ (\sigma_\text{sys}^{(i)})^2} + \left(\frac{\eta}{\sigma_\eta}\right)^2\right],
\end{align}
which will be used to derive constraints in the next subsection.

We note that the critical uncertainty is that of the experimental measurement of the $2\nu\beta\beta$ decay rate and not that in theoretical calculation of the corresponding nuclear matrix elements. This is because both the SM $2\nu\beta\beta$ decay and the one involving a sterile neutrino ($\nu N\beta\beta$) have the same nuclear matrix element and depend e.g. on the axial coupling strength $g_A$ in the same way, at least to a very good approximation as detailed below. Thus, while the individual decay rates have a large theoretical uncertainty, e.g. considering a range of $0.7 \lesssim g_A \lesssim 1.27$, their ratio is largely unaffected and one may use the experimental measurement to set the overall scale.

The heavier mass of the sterile neutrino does influence the energy denominators in Eq.~\eqref{eq:e-denominators} which changes the matrix elements as a sub-leading effect. This mostly affects differential decay properties, such as the electron energy spectrum, but it is essentially negligible for the sterile neutrino case with a left-handed current. This is because the distinctive feature, the different energy threshold for the $\nu N\beta\beta$ case, is unaffected: its location is determined by kinematics and its shape is already smooth, $\propto (Q-m_N-E)^{7/2}$, with small corrections having no discernable effect within the experimental energy resolutions considered. In other words, there is no sharp threshold (as in single $\beta$ decay) which is in danger of being washed out due to corrections.

The same procedure can be applied to place upper limits on the RH current couplings $|\epsilon_{LR}|^2$ and $|\epsilon_{RR}|^2$. As seen in the previous sections, the RH current modifies the total kinetic energy distribution to
\begin{align}
\label{eq:total_dist_RH}
	\frac{d\Gamma^{2\nu}(\boldsymbol{\xi})}{dE_K}
	 = \frac{d\Gamma^{2\nu}_\text{SM}}{dE_K} 
	 + |\epsilon_{XR}|^2 \, \frac{d\Gamma^{2\nu}_N(m_N)}{dE_K}\,,
\end{align}
where the SM contribution is no longer reduced by the sterile neutrino mixing. The RH current also modifies the angular distribution to Eq. \eqref{eq:angular_dist} with the total rate $\Gamma^{2\nu}$ and the angular correlation factor $K^{2\nu}$ given in terms of SM and RH current contributions as
\begin{align}
\label{eq:angularvariables}
	\Gamma^{2\nu}(\boldsymbol{\xi}) 
	= A^{2\nu}_\text{SM} + A^{2\nu}_N(m_N)|\epsilon_{XR}|^2\,, \quad 
	K^{2\nu}(\boldsymbol{\xi}) 
	= \frac{B^{2\nu}_\text{SM} + B^{2\nu}_N(m_N)|\epsilon_{XR}|^2 }{A^{2\nu}_\text{SM} 
	+ A^{2\nu}_N(m_N)|\epsilon_{XR}|^2}\,.
\end{align}
Assuming $|\epsilon_{XR}|^2\ll 1$, $K^{2\nu}$ can be Taylor expanded as
\begin{align}
\label{eq:angularvariables2}
	K^{2\nu}(\boldsymbol{\xi}) \approx K^{2\nu}_\text{SM} + \alpha(m_N)|\epsilon_{XR}|^2\,,
\end{align}
where the SM contribution and RH current contributions, respectively, are
\begin{align}
\label{eq:angularvariables3}
	K^{2\nu}_\text{SM} = \frac{B^{2\nu}_\text{SM}}{A^{2\nu}_\text{SM}}\,, \quad  
	\alpha(m_N) = \frac{B^{2\nu}_N(m_N) - K^{2\nu}_\text{SM} A^{2\nu}_N(m_N)}{A^{2\nu}_\text{SM}}\,,
\end{align}
\begin{figure}[t!]
	\centering
	\includegraphics[width=0.7\textwidth]{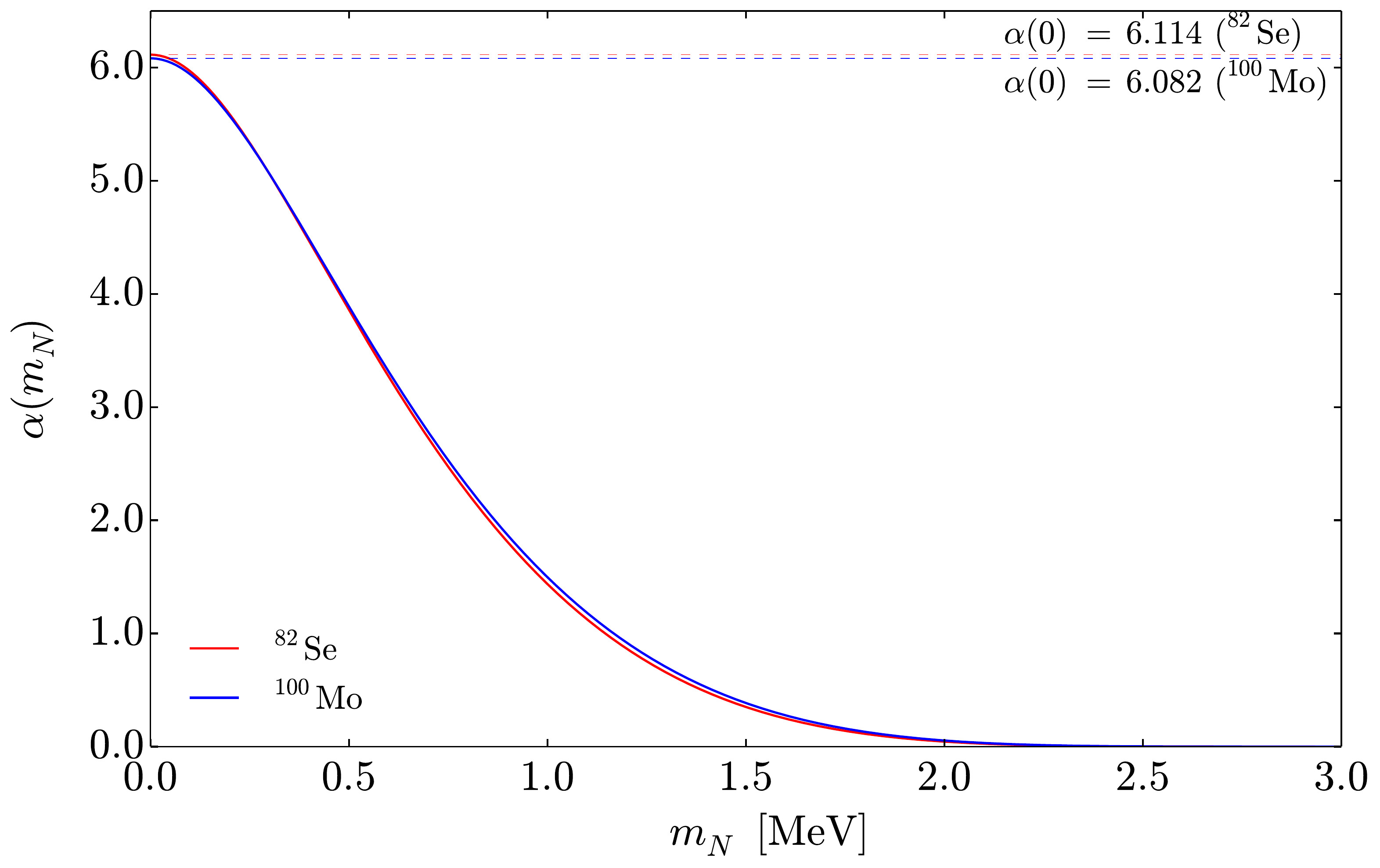}
	\caption{The approximate factor $\alpha(m_N)$ multiplying the RH current coupling $|\epsilon_{XR}|^2$ yielding the sterile neutrino contribution to the angular correlation factor $K^{2\nu}$ for $^{82}$Se (red) and $^{100}$Mo (blue).}
	\label{fig:alpha_plot}
\end{figure}
The SM values are $K^{2\nu}_\text{SM} = -0.627$ for $^{100}$Mo and $K^{2\nu}_\text{SM} = -0.631$ for $^{82}$Se (the isotopes of experiments that are sensitive to the angular correlation factor, NEMO-3 and SuperNEMO, respectively). The $\alpha(m_N)$ factors are plotted for $^{82}$Se (red) and $^{100}$Mo (blue) in Fig.~\ref{fig:alpha_plot}, which also indicates the values at $m_N = 0$. The factor $\alpha(m_N)$ is positive, indicating a change of the angular distribution away from the back-to-back configuration of electrons in the SM $V-A$ case. It is maximal for $m_N = 0$ and is suppressed to zero as $m_N$ approaches the $Q$ value.

Using the measured total kinetic energy distributions from all $2\nu\beta\beta$ decay experiments, the $\boldsymbol{\xi}=(m_N,|\epsilon_{XR}|^2)$ parameter space can be constrained in the same was as $(m_N, |V_{eN}|^2)$ described above, i.e. using the test-statistic in Eq.~\eqref{eq:finalteststatistic}. In addition, the experiments NEMO-3 and SuperNEMO will measure a certain number of events $N_\text{obs}^{(i)}$ distributed in bins of the cosine of the angle, $\cos\theta$. We can estimate the total number of signal plus background events $N^{(i)}_\text{exp}$ in each bin by integrating over the angular distribution Eq.~\eqref{eq:angular_dist}. We can then compute the test-statistic in Eq.~\eqref{eq:finalteststatistic} to put an additional constraint on the $\boldsymbol{\xi}$ parameter space.
	
\subsection{Results}

A selection of current and next generation $0\nu\beta\beta$ decay search experiments measuring the $2\nu\beta\beta$ decay of isotopes $^{76}$Ge, $^{82}$Se, $^{100}$Mo and $^{136}$Xe are shown in Tab.~\ref{tab:experiments}. Listed are the exposures, total number of events $N_\text{events}$, energy resolutions $\Delta E$ and estimates for the parameters $\sigma_\eta$ and $\sigma_f$ quantifying the uncertainties on the nuisance parameter $\eta$ and from other systematic effects, respectively. Values are taken from the list of references given for the experiments. For each experiment we make use of Eq.~\eqref{eq:finalteststatistic} to set an upper limit on the active-sterile mixing $|V_{eN}|^2$ as a function of the sterile neutrino mass $m_N$.
\begin{table}[b!]
	\setlength{\tabcolsep}{5.5pt}
	\begin{tabular}{cccccc}
		\hline\hline
		Isotope & Experiment &Exposure [$\mathrm{kg}\cdot\mathrm{y}$]& $N_{\text{events}}$ & $\Delta E$ [keV] & $(\sigma_\eta,\sigma_f)\,[\%]$ \\\hline
		\multirow{2}{*}{$^{76}$Ge}& GERDA II \cite{Agostini:2020xta} & $103.7$ & $3.63\times 10^4$  & $15$ & $(4.6, 1.9)$  \\
		& LEGEND \cite{Zsigmond:2020bfx} & $10^3\text{--}10^4$ & $10^5\text{--}10^6$ & $2.5$ & $(0.5, 0.5)$ \\\hline
		\multirow{2}{*}{$^{82}$Se}& CUPID-0 \cite{Azzolini:2019yib} & $9.95$ & $5.8\times 10^3$ & $50$ & $(1.5, 1.0)$  \\
		&  SuperNEMO \cite{Waters:2017wzp}  & $10^2\text{--}10^3$ & $10^4\text{--}10^5$ & $50$ & $(0.5, 0.5)$ \\\hline
		\multirow{3}{*}{$^{100}$Mo}& NEMO-3 \cite{NEMO-3:2019gwo} & $34.3$ & $4.95\times 10^5$ & $100$ & $(5.4, 1.8)$ \\
		& CUPID-Mo \cite{Armengaud:2019rll} & $0.116$ & $3.9\times 10^4$ & 20 & $(1.4, 0.5)$ \\
		& CUPID \cite{CUPIDInterestGroup:2019inu} & $10^2\text{--}10^3$ & $10^6\text{--}10^7$ & $5$ & $(0.5, 0.5)$ \\\hline
		\multirow{2}{*}{$^{136}$Xe}& KamLAND-Zen \cite{KamLAND-Zen:2019imh} & $126.3$ & $9.83\times 10^4$ & $50$ & $(3.1, 0.3)$ \\
		& DARWIN \cite{Agostini:2020adk} &$(2\text{--}5)\times 10^4$& $10^6\text{--}10^7$ & $5$ & $(0.5, 0.5)$ \\\hline\hline
	\end{tabular}
	\caption{Current and next generation $0\nu\beta\beta$ decay search experiments measuring the $2\nu\beta\beta$ decay spectrum of the isotopes considered in this work. Shown are the current and forecasted exposures, total number of events $N_\text{events}$, energy resolutions $\Delta E$ and parameters $(\sigma_\eta$, $\sigma_f)$ estimating the effect of systematic errors on the log-likelihood function.}
	\label{tab:experiments}
\end{table}

\begin{figure}[t!]
	\centering
	\includegraphics[width=0.495\textwidth]{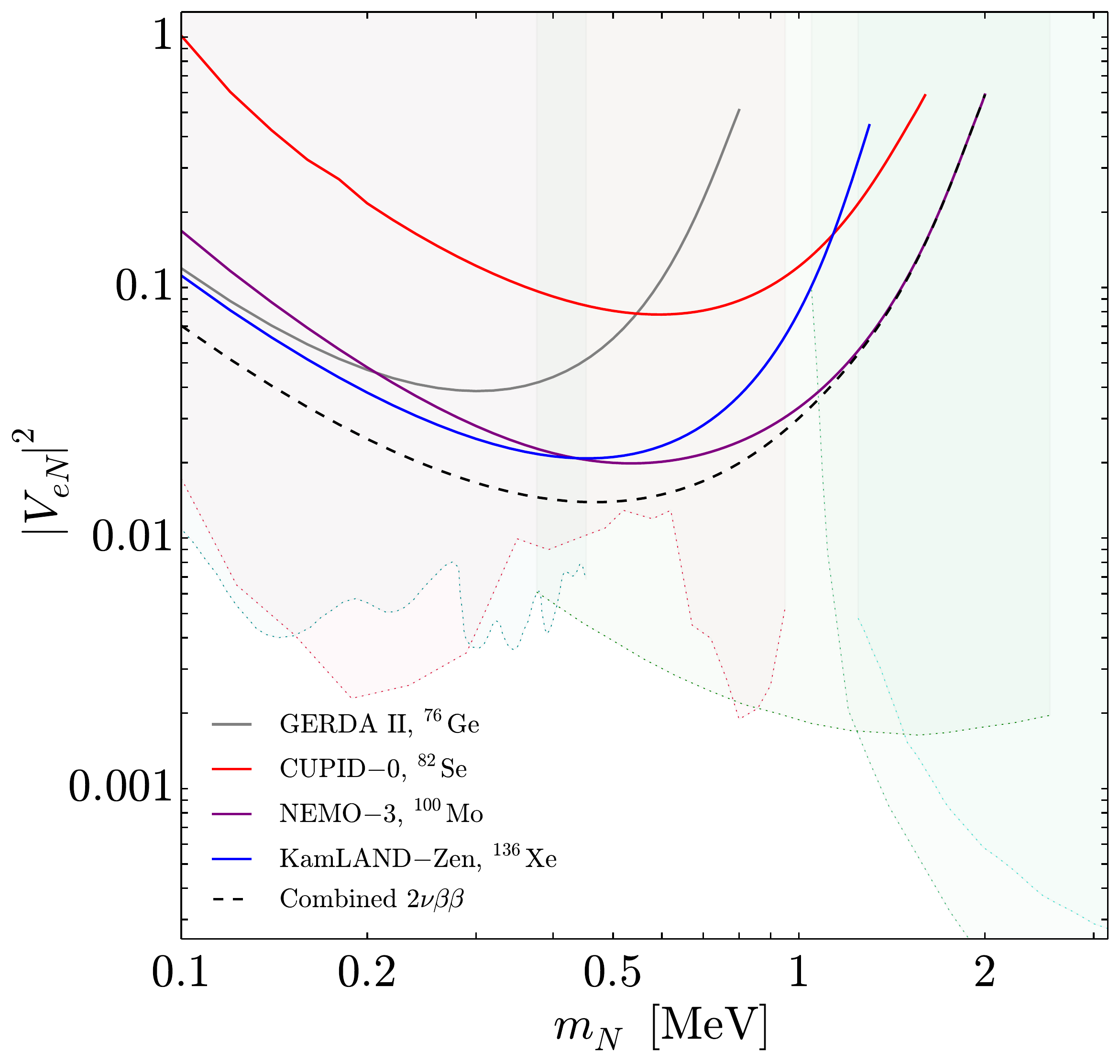}\hfill
	\includegraphics[width=0.495\textwidth]{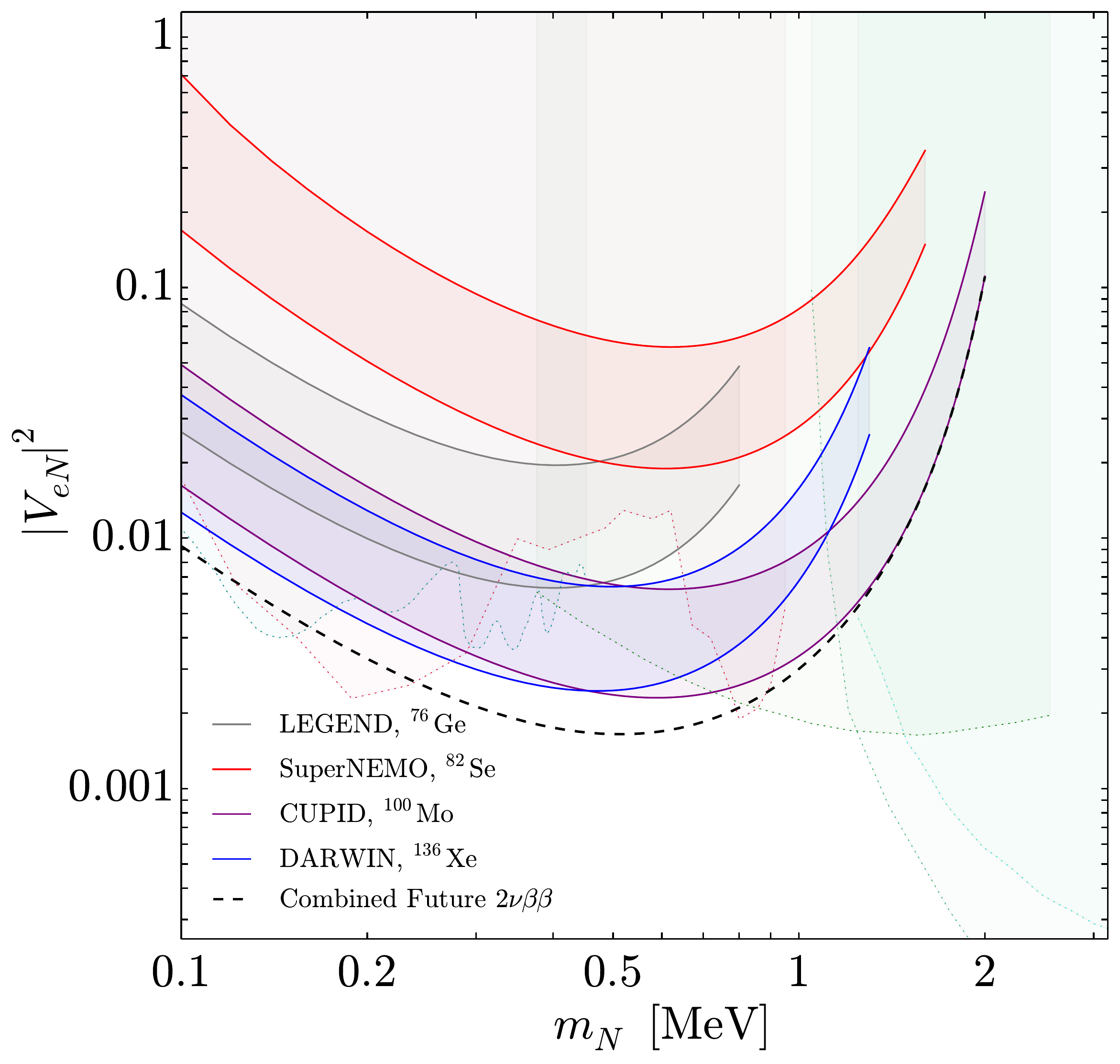}
	\caption{Upper limits and sensitivities at 90\% CL on the squared mixing $|V_{eN}|^2$ between the electron and sterile neutrino as a function of the sterile neutrino mass $m_N$ from $2\nu\beta\beta$ in current (left) and future (right) experiments. Shown are the individual constraints as indicated in the legend as well as a combined constraint (black dashed). The bands in the right plot correspond to the possible future exposures in Tab.~\ref{tab:experiments}. The  combined future sensitivity uses the maximum forecasted exposure of each experiment.}
	\label{fig:upper_limit_plot}
\end{figure}

Fig.~\ref{fig:upper_limit_plot}~(left) shows the 90\% CL upper limits derived from the current generation experiments GERDA II ($^{76}$Ge, grey), CUPID-0 ($^{82}$Se, red), NEMO-3 ($^{100}$Mo, purple) and KamLAND-Zen ($^{136}$Xe, blue). We also show a combined constraint (black dashed) found by summing the log-likelihoods of the experiments (each minimised with respect to a separate nuisance parameter $\eta$). It can be seen that the upper limits worsen for smaller and larger values of the sterile mass in the range $0.1~\text{MeV} < m_N < 3$~MeV, with the most stringent upper bound being found at $m_N$ similar to the peak energy of the associated spectrum. The constraints are compared to pre-existing constraints (shaded areas) from single beta decay experiments and sterile neutrino decays. While NEMO-3 and KamLAND-Zen provide the best individual constraints ($|V_{eN}|^2\lesssim 0.02$), they are not as competitive as previous limits. However, it is promising that $2\nu\beta\beta$ decay is more sensitive for sterile masses $0.3~\text{MeV} < m_N < 0.7$~MeV where existing constraints are less stringent.

Fig.~\ref{fig:upper_limit_plot}~(right) shows the corresponding sensitivities estimated for the next generation of $0\nu\beta\beta$ decay experiments. The forecasted range of exposures given by the collaborations are often one or two orders of magnitude larger than those of the current generation. We estimate the total number of events $N_\text{events}$ seen in future by multiplying the current values by the ratio of future to current exposures. Energy resolutions are taken from the references in Tab.~\ref{tab:experiments} and we assume an optimistic value of $\sigma_\eta \sim \sigma_f\sim 0.5\%$ for the systematic uncertainties. We compute the 90\% CL sensitivity for both the higher and lower forecasted number of events in Tab.~\ref{tab:experiments}, shown as bands for LEGEND ($^{76}$Ge, grey), SuperNEMO ($^{82}$Se, red), CUPID ($^{100}$Mo, purple) and DARWIN ($^{136}$Xe, blue). Also shown is the combined sensitivity (black dashed) using the largest predicted exposure of each experiment. For a given experiment the upper bounds exhibit the same improvement for sterile masses close to the maximum of the total differential decay rate. The most stringent upper limits come from CUPID and DARWIN, $|V_{eN}|^2\lesssim 2.5\times 10^{-3}$, which would exclude the currently unconstrained region in the $0.3~\text{MeV} < m_N < 0.7$~MeV range.

\begin{figure}[t!]
	\centering
	\includegraphics[width=0.6\textwidth]{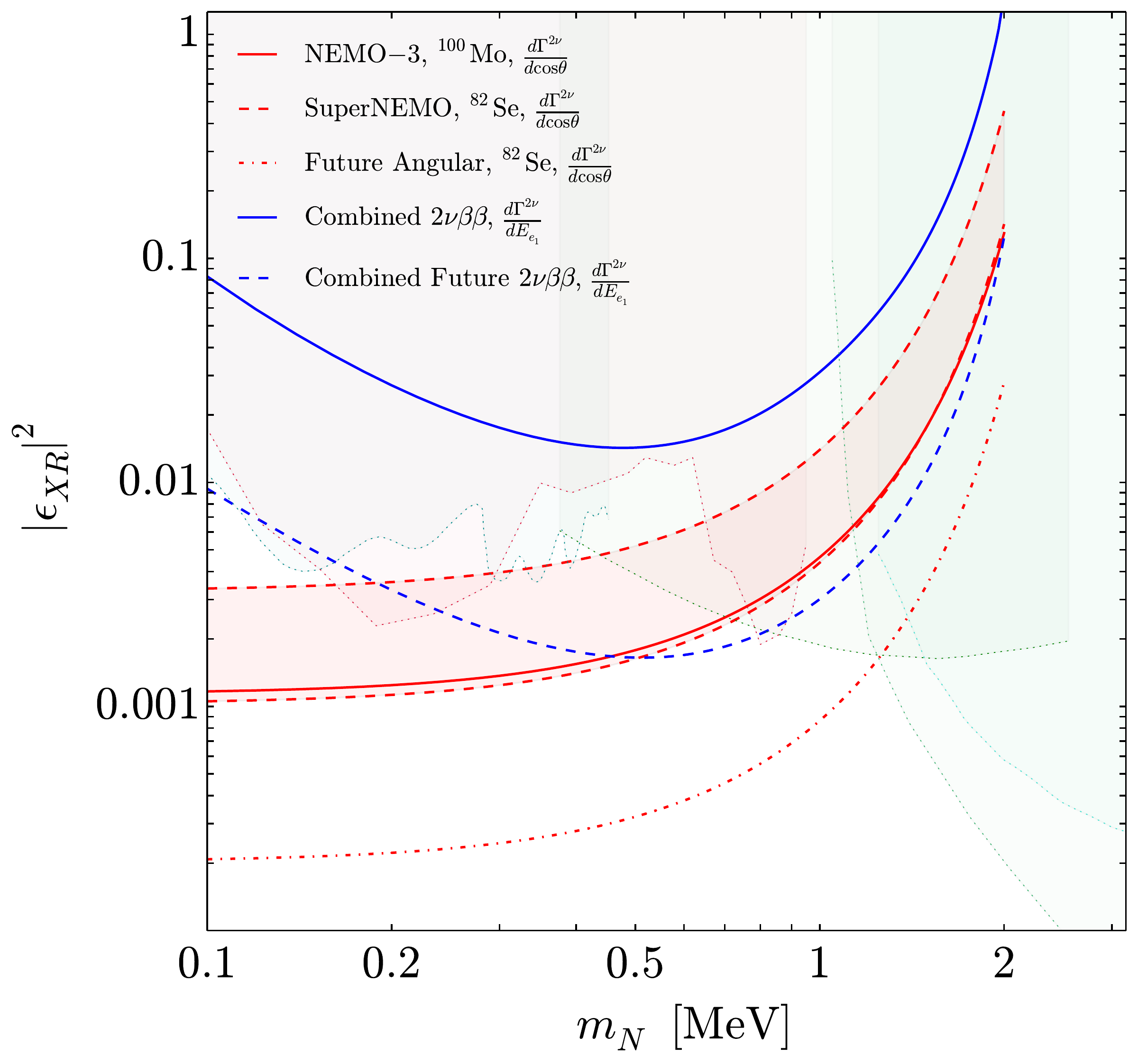}
	\caption{Current upper limits and future sensitivities at 90\% CL on the RH coupling $|\epsilon_{XR}|^2$ as a function of the sterile neutrinos mass $m_N$. The solid (dashed) blue line shows the combined constraint from current (future) $2\nu\beta\beta$ decay experiments measuring the total kinetic energy distribution. The solid red line is the upper limit derived from the angular distribution measurement of NEMO-3 ($^{100}$Mo). The dashed red band indicates the range of upper limits expected from the angular distribution measurement of SuperNEMO ($^{82}$Se). The dot-dashed red line shows the upper limit from a future $^{82}$Se experiment with an exposure of $10^7$ events.} 
	\label{fig:upper_limit_plot_RH}
\end{figure}
Likewise, we estimate the current limits and future sensitivity on the RH couplings $|\epsilon_{LR}|^2$ and $|\epsilon_{RR}|^2$ from measuring the $2\nu\beta\beta$ decay energy distribution and angular correlation. In Fig.~\ref{fig:upper_limit_plot_RH} we plot the upper limits at 90\% CL on $|\epsilon_{LR}|^2$ and $|\epsilon_{RR}|^2$ as a function of the sterile neutrino mass $m_N$. The blue solid line is the combined constraint from current $2\nu\beta\beta$ decay experiments using the total kinetic energy distribution, while the red solid line is the upper limit derived from the angular distribution measurement of NEMO-3 ($^{100}$Mo). The blue dashed line is the combined sensitivity from future $2\nu\beta\beta$ decay experiments, while the red dashed band indicates the sensitivity range from the angular distribution measurement of SuperNEMO ($^{82}$Se). The latter does not improve over the current limit as SuperNEMO is not expected to have a significantly increased exposure compared to NEMO-3, see Tab.~\ref{tab:experiments}. We therefore also indicate the sensitivity of a hypothetical $^{82}$Se angular measurement with an exposure of $10^7$ events (red dot dashed).

Due to the different total kinetic energy distribution for the RH current in Eq.~\eqref{eq:total_dist_RH} (no suppression of the SM rate), the combined constraints on $|\epsilon_{LR}|^2$ and $|\epsilon_{RR}|^2$ (dashed lines) are slightly weaker than the equivalent constraints on $|V_{eN}|^2$. The constraints from the NEMO-3 angular distribution are generally better, tending to a constant upper bound $|\epsilon_{XR}|^2 \lesssim 10^{-3}$ for $m_N \lesssim 0.2$~MeV. This roughly agrees with the result $\epsilon_{XR} < 2.7\times 10^{-2}$ in the massless case found in Ref.~\cite{Deppisch:2020mxv}.

\section{Conclusions}
\label{sec:conclusions}

\begin{figure}[t!]
	\centering
	\includegraphics[width=0.9\textwidth]{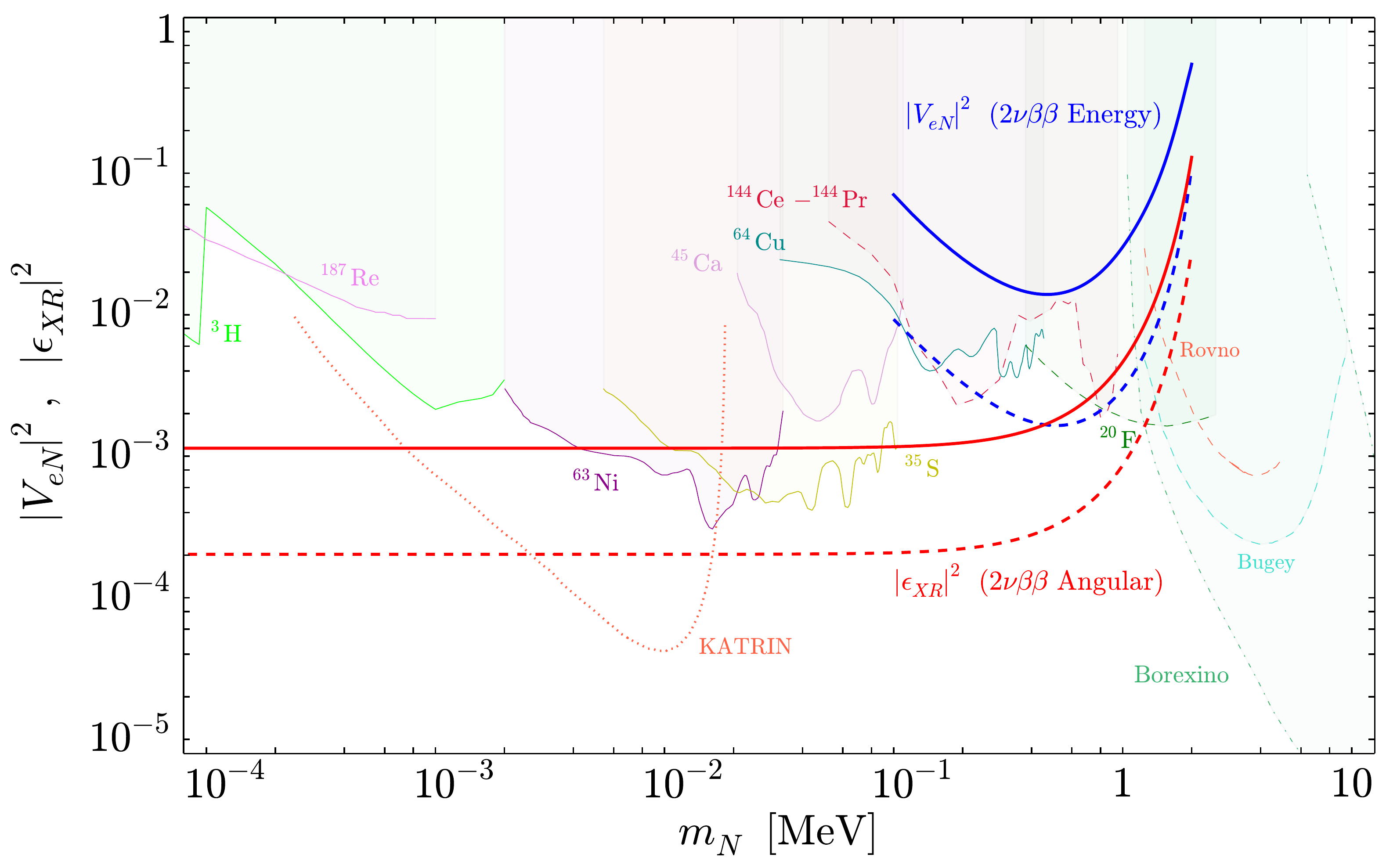}
	\caption{Current upper limits (solid blue) and future sensitivities (dashed blue) on the mixing strength $|V_{eN}|^2$ between the electron and sterile neutrino as a function of the sterile mass $m_N$. Likewise, the red curves give the current limit and future sensitivity on the RH coupling $|\epsilon_{XR}|^2$ using a measurement of the angular distribution in $2\nu\beta\beta$ decay. The shaded regions are excluded by existing searches in single beta decay and sterile decays in reactor and solar neutrino oscillation experiments.}
	\label{fig:comparison_plot}
\end{figure}

Measuring the kinematic endpoint in single beta decay is arguably the cleanest means to determine the absolute neutrino masses in a model-independent fashion. For the light active neutrinos in the SM, the most promising isotope for this is tritium ($^{3}\mathrm{H}$) and its beta decay is currently measured in the KATRIN experiment \cite{Aker:2019uuj} as well as the future Project~8 \cite{Esfahani:2017dmu} and CRESDA \cite{Saakyan:2020cresda} efforts. The same method can be applied to search for sterile neutrinos, not only in Tritium but in a host of beta decay isotopes where masses smaller than the respective $Q$ value of the decay can be probed. The limits on the active-sterile mixing strength $|V_{eN}|^2$ from such searches are summarised in Fig.~\ref{fig:comparison_plot}. They are comparatively weak, of the order $|V_{eN}| \lesssim 2\times 10^{-2} - 2\times 10^{-3}$, in the sterile neutrino mass range $ 0.1 \,\mathrm{MeV} < m_{N} < 1$~MeV. 

In this work, we have analysed the prospects to search for sterile neutrinos using the same principle in $2\nu\beta\beta$ decay. If one of the two neutrinos emitted in the process is a heavier, sterile neutrino it will likewise affect the distribution with respect to the kinetic energy of the two electrons observed in the decay: the kinematic endpoint is shifted to lower values depending on the sterile neutrino mass and the active-sterile mixing will reduce the usual SM contribution. This is expected to be challenging because of the very long $2\nu\beta\beta$ decay half lives and small rates compared to single beta decay. Nevertheless, future searches for the lepton number violating $0\nu\beta\beta$ decay will push the envelope in terms of exposure and allow measuring $2\nu\beta\beta$ decay with up to $10^7$ events. These data can then be used to probe exotic physics with $2\nu\beta\beta$ decay in its own right. Apart from sterile neutrino searches, other examples include exotic neutrino self interactions \cite{Deppisch:2020sqh} and RH leptonic currents \cite{Deppisch:2020mxv}. We have extended the latter analysis here to consider a RH $V+A$ current for a sterile neutrino rather than the SM electron neutrino. As in Ref. \cite{Deppisch:2020mxv}, this gives rise to an anomalous angular distribution of the electrons in $2\nu\beta\beta$ decay.

To summarise the sensitivity we compare in Fig.~\ref{fig:comparison_plot} the current limits on $|V_{eN}|^2$ from existing $2\nu\beta\beta$ decay (solid blue) to constraints from single beta decays and sterile neutrino decays over a wider range of masses, $100~\text{eV} < m_N < 10$~MeV. The blue curve uses the combined constraints from measurements of $2\nu\beta\beta$ decay electron energies. The red curve shows the current constraint on the effective RH coupling $|\epsilon_{XR}|^2$ using the NEMO-3 angular distribution measurement. The dashed curves indicate the corresponding future sensitivities. At lower masses both the current and future upper limits on $|V_{eN}|^2$ cannot compete with existing constraints from $^{64}$Cu and $^{144}$Ce$-^{144}$Pr beta decays. At higher masses they are also less stringent than constraints from Borexino, Bugey and Rovno. It is the $0.3~\text{MeV} < m_N < 0.7$~MeV range where $2\nu\beta\beta$ decay can provide competitive constraints in the future, though we expect that similar improvements from $^{20}$F and $^{144}$Ce$-^{144}$Pr beta decays are also possible. The constraints on the RH coupling $|\epsilon_{XR}|^2$ using an angular distribution measurement in $2\nu\beta\beta$ decay is most sensitive for light sterile neutrino masses $m_N \lesssim 0.1$~MeV as the effect is phase space suppressed otherwise. We note, though, that the limits from single beta decays and the other processes shown strictly speaking apply to $|V_{eN}|^2$ only and need to be re-evaluated for a heavy neutrino coupling through a RH current. 

Our analysis demonstrates that $2\nu\beta\beta$ decay can be used to search for sterile neutrinos with masses lighter than $m_{N}\sim 1$~MeV. While current searches are not competitive with limits from single beta decays, future searches will have a much more increased statistics where effects of new physics can be tested. While sterile neutrinos in this mass range are also heavily constrained from astrophysical measurements and cosmological considerations, it is important to  improve our understanding using all available data.

\section*{Acknowledgements}
The authors would like to thank Alexander Derbin for useful discussions on the constraints from single beta decay. The authors would also like to thank Matteo Agostini, Elisabetta Bossio, Alejandro Ibarra and Xabier Marcano for useful discussions on the revision of the manuscript. F. F. D. and P. D. B. acknowledge support from the UK Science and Technology Facilities Council (STFC) via a Consolidated Grant (Reference ST/P00072X/1). F\v{S} acknowledges support by the VEGA Grant Agency of the Slovak Republic under Contract No. 1/0607/20 and by the Ministry of Education, Youth and Sports of the Czech Republic under
the INAFYM Grant No. CZ.02.1.01/0.0/0.0/16\_019/0000766.

\bibliographystyle{JHEP}
\bibliography{References}
\end{document}